\documentclass[a4paper,11pt]{article}
\usepackage{jheppub}

\usepackage{mathtools}

\DeclarePairedDelimiter\ket{\lvert}{\rangle}
\DeclarePairedDelimiterX\braket[2]{\langle}{\rangle}{#1\,\delimsize\vert\,\mathopen{}#2}

\title{\boldmath Study of Scalar Non Standard Interaction at Protvino to Super-ORCA experiment}

\author{Dinesh Kumar Singha$^{1}$,}
\author{Rudra Majhi$^{1,2}$,}
\author{Lipsarani Panda$^{3}$,}
\author{Monojit Ghosh$^{4}$}
\author{and Rukmani Mohanta$^{1}$}

\affiliation{$^1$\,School of Physics, University of Hyderabad, Hyderabad - 500046, India} 
\affiliation{$^2$\,Department of Physics, Nabarangpur College, Nabarangpur - 764063, Odisha, India}
\affiliation{$^3$\,School of Physical Sciences, National Institute of Science Education and Research,  An OCC of Homi Bhabha National Institute, Bhubaneswar  752050,  Odisha, India }
\affiliation{$^4$\,Center of Excellence for Advanced Materials and Sensing Devices, Ruder Bo\v{s}kovi\'c Institute, 10000 Zagreb, Croatia}

\emailAdd{dinesh.sin.187@gmail.com, rudra.majhi95@gmail.com, lipsarani.panda@niser.ac.in, mghosh@irb.hr, rmsp@uohyd.ac.in}

\abstract{In this paper we have studied the phenomenon of non-standard interaction mediated by a scalar field (SNSI) in the context of P2SO experiment and compared its sensitivity with DUNE. In particular, we have studied the capability of these two experiments to put bounds on the diagonal SNSI parameters i.e., $\eta_{ee}$, $\eta_{\mu\mu}$ and $\eta_{\tau\tau}$ and studied the impact of these parameters on the determination of neutrino mass ordering, octant of $\theta_{23}$ and CP violation (CPV). In our analysis we find that, the parameter $\Delta m^2_{31}$ has a non-trivial role if one wants estimate the bounds on $\eta_{\mu\mu}$ and $\eta_{\tau\tau}$ assuming SNSI does not exist in nature. Our results show that sensitivity of P2SO and DUNE to constraint $\eta_{\mu\mu}$ and $\eta_{\tau\tau}$ are similar whereas the sensitivity of DUNE is slightly better for $\eta_{ee}$. We find that the mass ordering and CPV sensitivities are mostly affected by $\eta_{ee}$ compared to $\eta_{\mu \mu}$ and $\eta_{\tau \tau}$ if one assumes SNSI exists in nature. On the other hand, octant sensitivity is mostly affected by $\eta_{\mu \mu}$ and $\eta_{\tau \tau}$. These sensitivities can be either higher or lower than the standard three flavour scenario depending on the relative sign of the SNSI parameters. Regarding the precision of atmospheric mixing parameters, we find that the precision of $\theta_{23}$ deteriorates significantly in the presence of $\eta_{\mu\mu}$ and $\eta_{\tau\tau}$. 

}

\makeatletter
\gdef\@fpheader{}
\makeatother
\begin{document}
\maketitle
\flushbottom

\section{Introduction}
\label{sec:intro}

The Standard Model (SM) of particle physics~\cite{Weinberg:1967tq}, despite being one of the most successful theories, proven incomplete often. One such example is the discovery of neutrino oscillation which shows neutrinos have non-zero masses. To explain this phenomena one requires theories beyond SM (BSM)~\cite{Bilenky:2016pep}. In the past few decades, neutrino masses and mixing, along with other intriguing properties, have been unraveled by many dedicated neutrino experiments. Some of the neutrino oscillation parameters are already measured  very precisely within the standard three flavour scenario. However, still there are some unknowns and degeneracies to be explored~\cite{Esteban:2020cvm}. Apart from the standard three flavour scenario, neutrino oscillation experiments can be used to probe several new physics scenarios. The example of one such new physics scenarios is non-standard interactions (NSI).

NSI mediated by a vector field is a very popular topic in the context of neutrino oscillation~\cite{Farzan:2017xzy}. NSI mediated by a vector field can be of charge current (CC) type and neutral current (NC) type. The CC-NSI leads to modifications
in the production and detection of neutrinos, whereas NC-NSI appears as a matter potential term in neutrino oscillation. However, there can also be NSI mediated by a scalar field (SNSI). This SNSI acts as an Yukawa term in the Lagrangian and therefore modifies the neutrino masses. If we assume the correction to the neutrino mass matrix due to SNSI is Hermitian, then the neutrino oscillation parameter space is increased by three real diagonal parameters and three complex off-diagonal parameters. Study of SNSI in the context of neutrino oscillation experiment is very new and therefore, in recent times there are several studies exploring various phenomenological aspects of SNSI~\cite{Ge:2018uhz,Ge:2019tdi,Denton:2022pxt,Cordero:2022fwb,Gupta:2023wct,Medhi:2021wxj,Medhi:2022qmu,Medhi:2023ebi,Sarkar:2022ujy,Babu:2019iml}. In this paper, we will study SNSI in the context of Protvino to Super-ORCA (P2SO)  experiment~\cite{Akindinov:2019flp} and compare its sensitivity with the DUNE experiment~\cite{DUNE:2020ypp}. The future P2SO experiment will use a neutrino beam from Protvino in Russia. In this case, the neutrinos will be detected at the Super-ORCA detector at the KM3NeT facility~\cite{KM3Net:2016zxf} located at a distance of 2595 km from Protvino. For the upcoming DUNE experiment, the neutrinos produced at Fermilab will be detected at a distance of 1300 km. For our present study, we will consider only the diagonal parameters whereas the effect of the off-diagonal parameters will be studied in a separate work. In this work, our objective will be to study the capability of P2SO and DUNE to put bounds on the three diagonal SNSI parameters and study the effect of these parameters in the measurement of the standard oscillation parameters. Our work is the first to put the future expected upper bounds on the SNSI parameters and in doing so we will point out a non-trivial role of the atmospheric mass square difference.

The paper is organized as follows. In the next section we will provide the theoretical background of SNSI and present the oscillation probabilities as a function of neutrino energy for P2SO. After that, we will briefly describe the experimental and simulation details used in our numerical analysis. Then we will proceed to present our numerical results. Finally, we will summarize our main findings and  conclude.

\section{Neutrino oscillation in presence of scalar NSI}
\label{bg}

The non-standard interaction between the neutrinos $\nu$ and the fermions $f$, mediated by a scalar field $\phi$ can be represented by the Feynman diagram shown in Fig. 1.
\begin{figure}[htb!]
    \centering
     \includegraphics[scale=0.6]{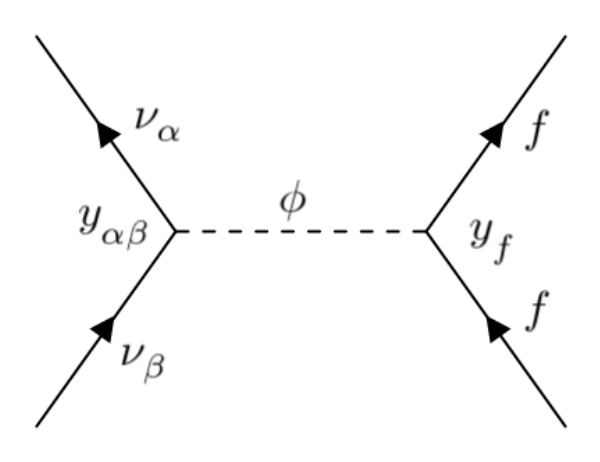}
    \caption{Feynmann diagram contributing to scalar NSI.}
    \label{feyn}
\end{figure}

In this case the effective Lagrangian can be written as,
\begin{eqnarray}
 \mathcal{L}_{\rm eff} = \frac{y_f y_{\alpha \beta}}{m_\phi^2} (\overline{\nu}_\alpha\nu_\beta)(\bar{f}{f}),
 \label{lag}
\end{eqnarray}
where $y$’s are the Yukawa couplings and $m_\phi$ is the mass of the scalar mediator. The Dirac equation in the presence of SNSI can be written as:
\begin{eqnarray}
\overline{\nu}_\beta \left[i \partial_\mu \gamma^\mu + \left(M_{\beta \alpha} + \frac{\sum_f N_f y_f y_{\alpha \beta}}{m_\phi^2}\right)\right] \nu_\alpha = 0\;,
\label{dir}
\end{eqnarray}
where $M_{\beta \alpha}$ is the Dirac mass matrix of the neutrinos and $N_f$ is the number density of fermion $f$. Therefore, we see that the effect of SNSI manifest as a correction term to the neutrino mass matrix. This correction can be parameterized as
\begin{eqnarray}
\delta M =  \sqrt{|\Delta m^2_{31}|}
\begin{pmatrix}
\eta_{ee} & \eta_{e\mu} & \eta_{e\tau}\\
\eta_{\mu e} & \eta_{\mu \mu} & \eta_{\mu \tau} \\
\eta_{\tau e} & \eta_{\tau \mu} & \eta_{\tau\tau}
\label{eta}
\end{pmatrix}\;,
\end{eqnarray}
where we have chosen to scale the size of $\delta M$ relative to $\sqrt{|\Delta m^2_{31}|}$ to make the SNSI parameters,  i.e., $\eta$ dimensionless. Comparing Eqs.~\ref{dir} and \ref{eta}, one can write
\begin{eqnarray}
    \eta_{\alpha \beta} = \frac{1}{m_\phi^2 \sqrt{|\Delta m^2_{31}|}} \sum_f N_f y_f y_{\alpha \beta}\;.
\label{eta_d}
\end{eqnarray}
Here $\Delta m^2_{31} = m_3^2 - m_1^2$ is the atmospheric mass square difference. Note that the bounds of the Yukawa couplings can come from the neutrino electron scattering experiments~\cite{Dutta:2022fdt} and cosmology~\cite{Venzor:2020ova}. It was shown in Ref.~\cite{Dutta:2022fdt} that same values of the couplings satisfy a wide range of mediator mass. Therefore, there will not be a direct correlation between the bounds of the SNSI parameters obtained from scattering experiments. We will consider $\delta M$ to be Hermitian with $\eta_{\alpha \alpha}$ as real and $\eta_{\alpha \beta}$ with $\alpha \neq \beta$ as complex. As mentioned in the introduction, for this present work, we will  consider only the diagonal parameters. Here it is important to note that, the parameter $\eta$ in Eq.~\ref{eta_d} depends on the density \footnote{Note that NSI mediated by a vector field (VNSI) also depends on the matter density. In the appendix, we have added a section discussing the separation of VNSI from SNSI.}. Therefore, when comparing the values of $\eta$ for different experiments, one should be careful to scale its value according to the matter density profile which was considered in those experiments. However, in our case, as the matter densities relevant for P2SO and DUNE are very similar, we will safely ignore this and compare the values of $\eta$ between this experiments without any scaling. 

Now let us see how this $\delta M$ modifies the Hamiltonian of the neutrino oscillation. The Hamiltonian of neutrino oscillation in the flavour basis and in presence of scalar NSI can be written as
\begin{eqnarray}
  H = E_\nu + \frac{\mathcal{M} \mathcal{M}^\dagger}{2E_\nu} + V \;,
  \label{ham}
\end{eqnarray}
where $E_\nu$ is the energy of the neutrinos, $V = {\rm diag}(\sqrt{2}G_F N_e,0,0)$ is the standard matter potential with $G_F$ is the Fermi constant and $N_e$ is the electron number density. In this case, the term $\mathcal{M}$ becomes
\begin{eqnarray}
 \mathcal{M} &=& U~{\rm diag}(m_1, m_2, m_3)~U^\dagger + \delta M \nonumber\\
   &=& U~{\rm diag}\left(m_1, \sqrt{m_1^2 + \Delta m^2_{21}}, \sqrt{m_1^2 + \Delta m^2_{31}}\right)~U^\dagger + \delta M\;,\label{h} 
 \end{eqnarray}
where we have assumed normal ordering of the neutrino masses i.e., $m_3 \gg m_2 > m_1$. Here $\Delta m^2_{21} = m_2^2 - m_1^2$ is the solar mass squared difference and $U$ is the PMNS matrix having the parameters $\theta_{12}$, $\theta_{13}$, $\theta_{23}$ and $\delta_{\rm CP}$. Neutrino oscillation probabilities in presence of SNSI can be calculated by diagonalizing Eq.~\ref{ham}. It is interesting to note that, for SNSI, the neutrinos oscillation probabilities will depend on the absolute neutrino mass $m_1$. 

\begin{figure} [htb!]
    \centering
    \includegraphics[scale=0.9]{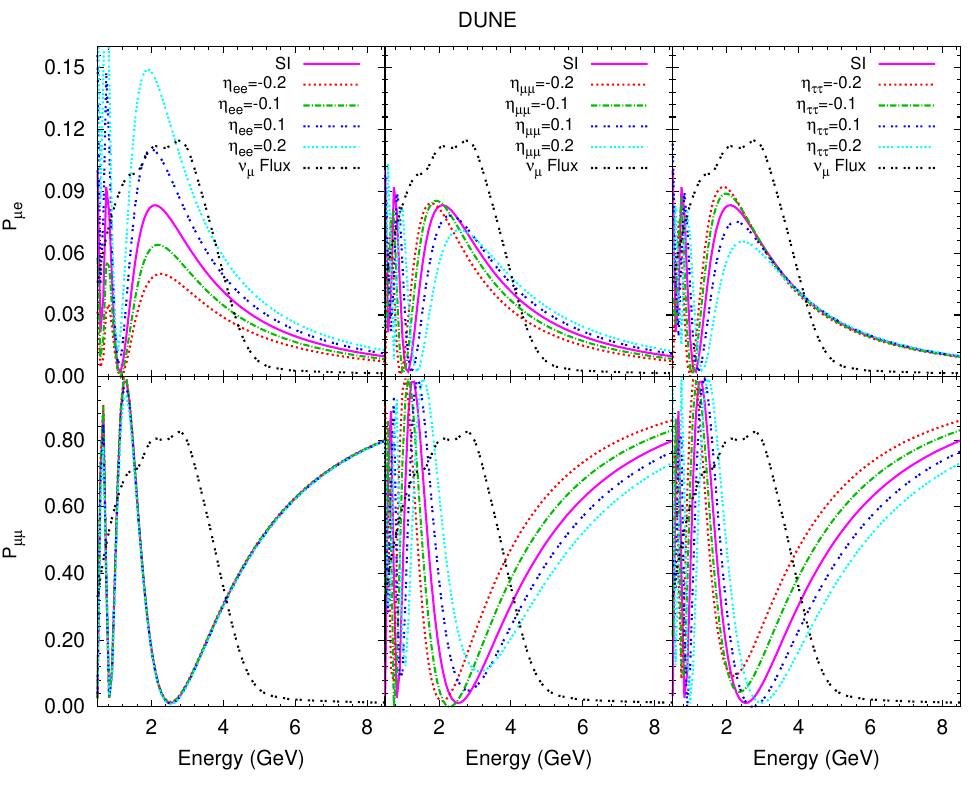}
    \caption{Electron neutrino appearance (disappearance) probability  as a function of neutrino energy for DUNE experiment in upper (lower) panel.}
    \label{fig:DUNE-prob}
\end{figure}

\begin{figure}[htb!]
    \centering
    \includegraphics[scale=0.9]{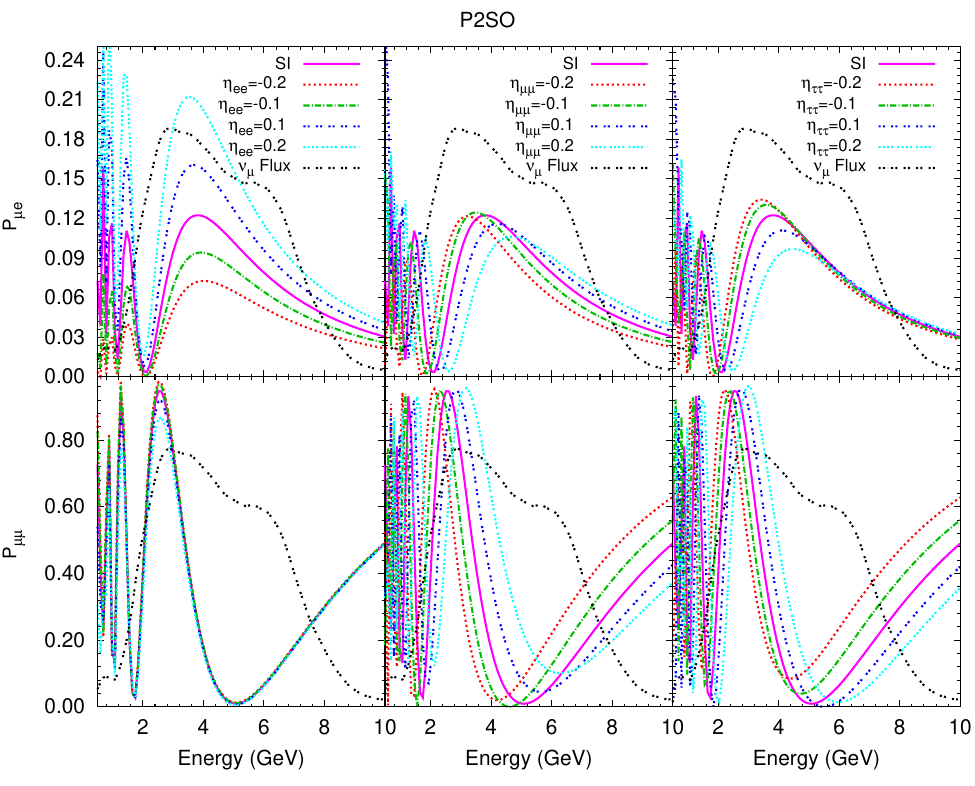}
    \caption{Electron neutrino appearance (disappearance) probability  as a function of energy for P2SO experiment in upper (lower) panel.}
    \label{fig:P2SO-prob}
\end{figure}

In order to calculate the neutrino oscillation probabilities in the presence of SNSI, we have modified the GLoBES~\cite{Huber:2004ka,Huber:2007ji} probability engine.
The results are shown in Fig.~\ref{fig:DUNE-prob} (for DUNE) and Fig.~\ref{fig:P2SO-prob} (for P2SO) for the neutrinos and assuming normal ordering of the neutrino masses. In this figure, we have considered one SNSI parameter at a time. In each figure, the top row is for the appearance channel i.e., $\nu_\mu \rightarrow \nu_e$, and the bottom row is for the disappearance channel i.e., $\nu_\mu \rightarrow \nu_\mu$. In each row, the left/middle/right panel is for $\eta_{ee}/\eta_{\mu\mu}/\eta_{\tau\tau}$. In each panel, we have shown the probabilities for the standard three flavour and probabilities in the presence of SNSI. For illustration, we have taken the values of the diagonal SNSI parameters to be $\pm 0.1$ and $\pm 0.2$. In addition, we have shown the energy dependence of the $\nu_\mu$ fluxes of each experiment. Regarding the values of the standard oscillation parameters, we have used the latest global analysis results of Nufit 5.2~\cite{Esteban:2020cvm} and they are given in Table~\ref{table_sparam}. To generate these figures, we have taken $m_1 = 10^{-5}$ eV.

\begin{table}[h] 
\centering
\begin{tabular}{|c|c|} \hline
Parameters            & True values $\pm$ $1\sigma$       \\ \hline
$\sin^2 \theta_{12}$  & $0.303^{+ 0.012}_{- 0.011}$      \\ 
$\sin^2 \theta_{13}$ & $0.02203^{+ 0.00056}_{- 0.00059}$                 \\ 
$\sin^2 \theta_{23} $ & $0.572^{+ 0.018}_{- 0.023}$                 \\ 
$\delta_{\rm CP}[^\circ] $  & $ 197^{+ 42}_{- 25}$         \\ 
$\Delta m^2_{21}$ [10$^{-5}$ eV$^2$]    & $7.41^{+ 0.21}_{- 0.20}$  \\ 
$\Delta m^2_{31}$ [10$^{-3}$ eV$^2$]   & $2.511^{+ 0.028}_{- 0.027}$    \\ 
 \hline
\end{tabular}
\caption{Oscillation parameter values  with their corresponding 1$\sigma$ errors considered in our analysis \cite{Esteban:2020cvm}.}
\label{table_sparam}
\end{table}   

From Fig.~\ref{fig:DUNE-prob} and Fig.~\ref{fig:P2SO-prob}, we observe that the sensitivity to $\eta_{ee}$ is expected to come from the appearance channel, whereas both the appearance and disappearance channels are sensitive to $\eta_{\mu\mu}$ and $\eta_{\tau\tau}$. Among the three parameters, the best sensitivity is expected for the parameter $\eta_{ee}$. We also see some interesting features regarding the positive and negative values of the SNSI parameters. The probabilities for a negative (positive) values of $\eta_{ee}$ are lower (higher) as compared to the probabilities in the standard case in the appearance channel. This is opposite in the case of $\eta_{\mu\mu}$ and $\eta_{\tau\tau}$. In the disappearance (appearance) channel, the position of the first oscillation minimum (maximum) is shifted in the opposite directions due to positive and negative values of $\eta_{\mu\mu}$ and $\eta_{\tau\tau}$. In addition, the probabilities at the first oscillation minimum shifts from zero for the positive (negative) values of $\eta_{\mu\mu}$ ($\eta_{\tau\tau}$) in the disappearance channel. 

Note that currently an analytical expression for the appearance channel probabilities involving the diagonal SNSI parameters does not exist in the literature and this derivation is very complicated in nature. We have attempted to derive an analytical expression for $\eta_{ee}$ which we discuss in the appendix. For disappearance channel,  in Ref.~\cite{Denton:2022pxt}, it was shown that at the leading order, the probability in the presence of SNSI can be written as
\begin{eqnarray}
P_{\mu\mu} = -\Delta_{13} (\eta_{\mu\mu} + \eta_{\tau\tau} + 2 |\eta_{\mu\tau}| \cos\phi_{\mu\tau}) \sin2\Delta_{13},
\end{eqnarray}
where $\Delta_{13} = \Delta m^2_{31} L/4E$, with $L$ being the baseline and $\phi_{\mu\tau}$ being the phase of the parameter $\eta_{\mu\tau}$. From the above equation we see that $\eta_{\mu\mu}$ and $\eta_{\tau\tau}$ appear with the same sign in the probability. This explains the shift of the oscillation minima for both $\eta_{\mu\mu}$ and $\eta_{\tau\tau}$ in the same directions for their positive/negative values. The shifting of the minima from zero for these parameters due to the opposite signs of these parameters (i.e., positive values for $\eta_{\mu\mu}$ and negative values for $\eta_{\tau\tau}$) might have caused due to some sub-leading terms in the probability. From the above equation, we also note that the disappearance channel probability does not depend upon the parameter $\eta_{ee}$ at the leading order. This confirms our numerical observation which shows that the disappearance probabilities do not change with the variation of the parameter $\eta_{ee}$ for both P2SO and DUNE.

\section{Experimental Details}

For simulating the long-baseline experiment P2SO, we use the same configuration as used in Ref.~\cite{Singha:2022btw}.  The Protvino accelerator with a 1.5 km-diameter U-70 synchrotron will generate 450 KW beam to produce 4 $\times 10^{20}$ protons on target (POT) per year.
The neutrinos will be detected at the Super-ORCA detector which will have a fiducial volume in megaton (Mt) scale. We have considered a total run-time of six years, divided into three years in neutrino mode and three years in antineutrino mode.

For DUNE, we have used the official GLoBES files corresponding to the technical design report~\cite{DUNE:2021cuw}. For DUNE, the detector will be a 40 kt liquid argon time-projection chamber. The beam power in this case will be 1.2 MW. The total run-time for DUNE which we considered in our work is 13 years comprising of 6.5 years in neutrino mode and 6.5 years in antineutrino mode. This corresponds to $1.1 \times 10^{21}$ POT per year. 

For the estimation of the sensitivity, we use the Poisson log-likelihood and assume that it is $\chi^2$-distributed:
\begin{equation}
 \chi^2_{{\rm stat}} = 2 \sum_{i=1}^n \bigg[ N^{{\rm test}}_i - N^{{\rm true}}_i - N^{{\rm true}}_i \log\bigg(\frac{N^{{\rm test}}_i}{N^{{\rm true}}_i}\bigg) \bigg]\,,
\end{equation}
where $N^{{\rm test}}$ and $N^{{\rm true}}$ are the number of events in the test and true spectra respectively, and $i$ is the number of energy bins. The systematic is incorporated by the method of pull \cite{Fogli:2002pt,Huber:2002mx}. For systematic uncertainties, we have considered the overall normalization and shape errors corresponding to signal and background. We list the values of systematic errors for P2SO and DUNE in Table \ref{table_sys}. It should be noted that the DUNE GLoBES file contains no shape error. We show all our results for the normal hierarchy of the neutrino masses with $m_1 = 10^{-5}$ eV unless otherwise specified. The true values of the parameters that we use in our analysis are given in Table~\ref{table_sparam}. In all our results, we will consider one SNSI parameter at a time. 

\begin{table} 
\centering
\begin{tabular}{|c|c|c|} \hline
Systematics     & P2SO          & DUNE  \\ \hline
Sg-norm $\nu_{e}$   & 5$\%$   & 2$\%$      \\ 
Sg-norm $\nu_{\mu}$    & 5$\%$            & 5$\%$ \\ 
Bg-norm    & 12$\%$     & 5$\%$ to 20$\%$\\ 
Sg-shape      & 11$\%$     & -\\ 
Bg-shape     & 4\% to 11$\%$       & - \\ 
\hline
\end{tabular}
\caption{The values of systematic errors that we considered in our analysis. ``norm" stands for normalization error, ``Sg" stands for signal and ``Bg" stands for background.}
\label{table_sys}
\end{table}  

\section{Results}

\subsection{Bounds on the SNSI parameters}

\begin{figure}[htb!]
    \centering

\includegraphics[scale=0.275]{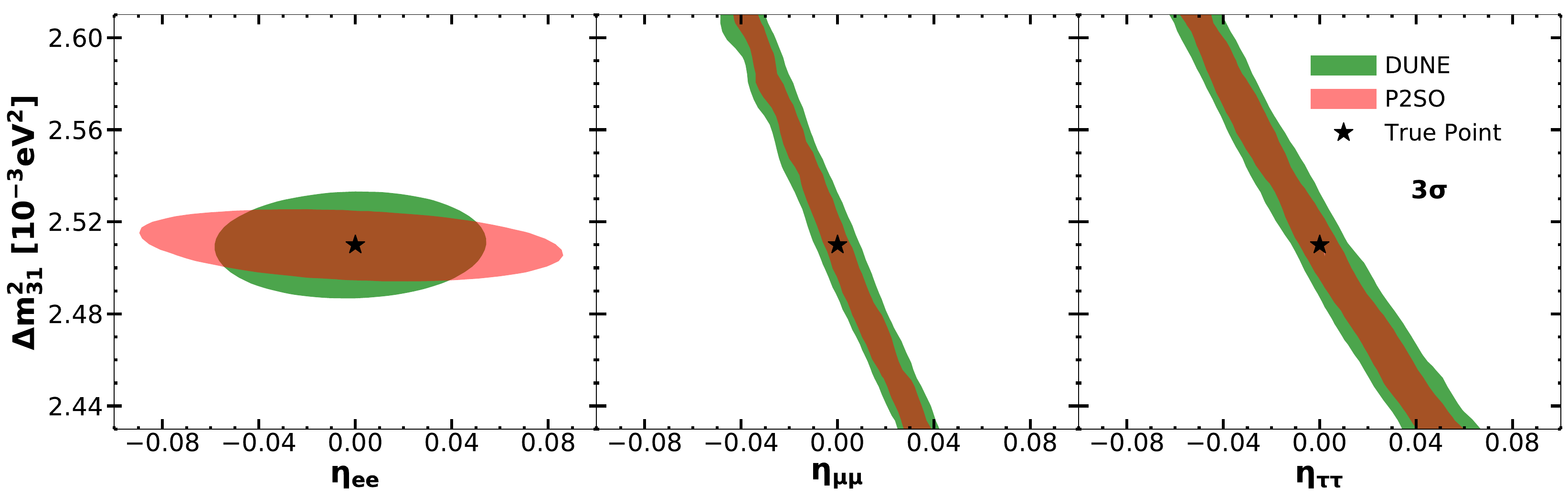}

    \caption{Allowed values of $\Delta m^2_{31}$ at $3 \sigma$ C.L. when SNSI is fitted in the theory with the standard three flavour scenario in the data.  }
    \label{fig:SNSI_fit}
\end{figure}

First, let us try to see the capabilities of P2SO and DUNE to constrain the diagonal SNSI parameters. Usually this is done by taking the standard three flavour scenario in the true spectrum of the $\chi^2$, taking SNSI in the test spectrum of the $\chi^2$ and plotting the $\chi^2$ against the SNSI parameters. Before presenting these results, it is important to understand the role of $\Delta m^2_{31}$ in putting the constraints on the SNSI parameters. For this, in Fig.~\ref{fig:SNSI_fit}, we have plotted $3 \sigma$ contour in the $\eta$ (test) - $\Delta m^2_{31}$ (test) plane, taking the standard three flavour scenario in true. In this figure, the left/middle/right panel is for $\eta_{ee}/\eta_{\mu\mu}/\eta_{\tau\tau}$. In generating this figure, all the other parameters  that are not shown (except $\delta_{\rm CP}$) are minimized randomly using their $1 \sigma$ error as priors as listed in Tab.~\ref{table_sparam}. The parameter $\delta_{\rm CP}$ is minimized without any prior. The green (red) shaded allowed region is for DUNE (P2SO). In these panels, the range in y-axis corresponds to the current $3 \sigma$ range of $\Delta m^2_{31}$ according to Nufit 5.2. 
From the figure we see that, for the parameter $\eta_{ee}$, the standard scenario cannot be fitted with SNSI having a value of $\Delta m^2_{31}$ lying outside the current $3 \sigma$ range. This is because in this case, the contours for both DUNE and P2SO are not extended beyond the plotted y-axis values of $\Delta m^2_{31}$. But for $\eta_{\mu\mu}$ and $\eta_{\tau\tau}$, the standard scenario can also be fitted with SNSI having a value of $\Delta m^2_{31}$ lying outside its current $3 \sigma$ allowed range. This can be understood by observing that for these two parameters, the contours are getting extended beyond the y-axis ranges of $\Delta m^2_{31}$. This implies that, when we calculate the bounds for $\eta_{\mu\mu}$ and $\eta_{\tau\tau}$, the $\chi^2$ minimum can correspond to a value of $\Delta m^2_{31}$ beyond the current $3 \sigma$ range\footnote{We have explicitly checked that this situation does not occur with any other parameter.}. Therefore, it is very important to take extra care when minimizing this parameter at the time of calculating the bounds of $\eta_{\mu\mu}$ and $\eta_{\tau\tau}$. If we minimize this parameter taking its $1 \sigma$ error as a prior, then there is a chance that the $\chi^2$ minimizer will take a local minima near to its true value. On the other hand, if we minimize this parameter without any prior, then the $\chi^2$ minimum will occur beyond its allowed $3 \sigma$ values. To avoid this situation, we will minimize this parameter by the method of systematic sampling rather than random sampling i.e., vary this parameter within its current $3 \sigma$ allowed range and then choose the $\chi^2$ minimum. Here it is important to note that the current $3 \sigma$ range of $\Delta m^2_{31}$ was calculated using the standard three flavour scenario. However, if this parameter is measured considering SNSI in theory, the allowed $3\sigma$ region may get extended accommodating larger values of this parameter\footnote{It will be very intriguing to see what happens to the measurement of oscillation parameters especially $\Delta m^2_{31}$ with respect to the current data if one assumes SNSI exists in Nature. However, this is beyond the scope of this present work.}. To incorporate this situation, one may consider to minimize this parameter with a flat prior. Since we do not know if that is the case, we decided to vary $\Delta m^2_{31}$ within its current $3 \sigma$ range.

\begin{figure}[htb!]
    \centering
\includegraphics[scale=0.355]{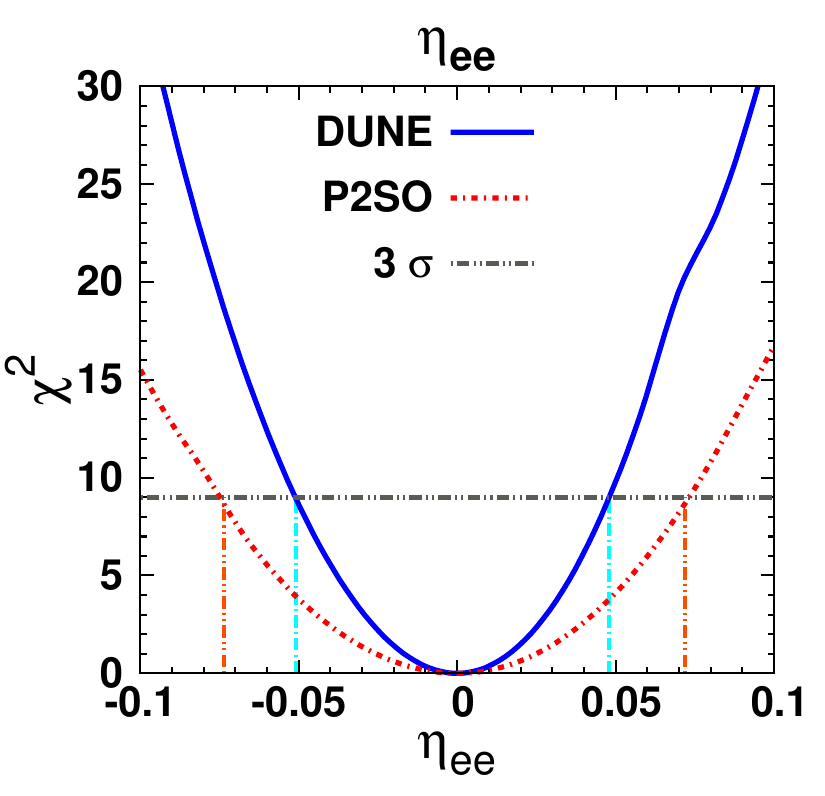}
\includegraphics[scale=0.355]{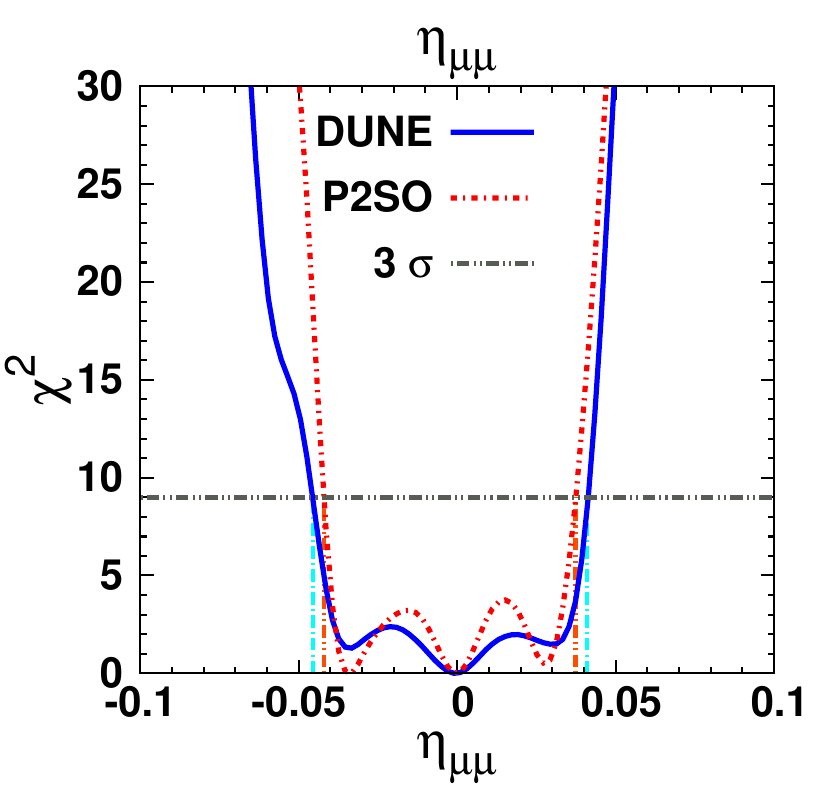}
\includegraphics[scale=0.355]{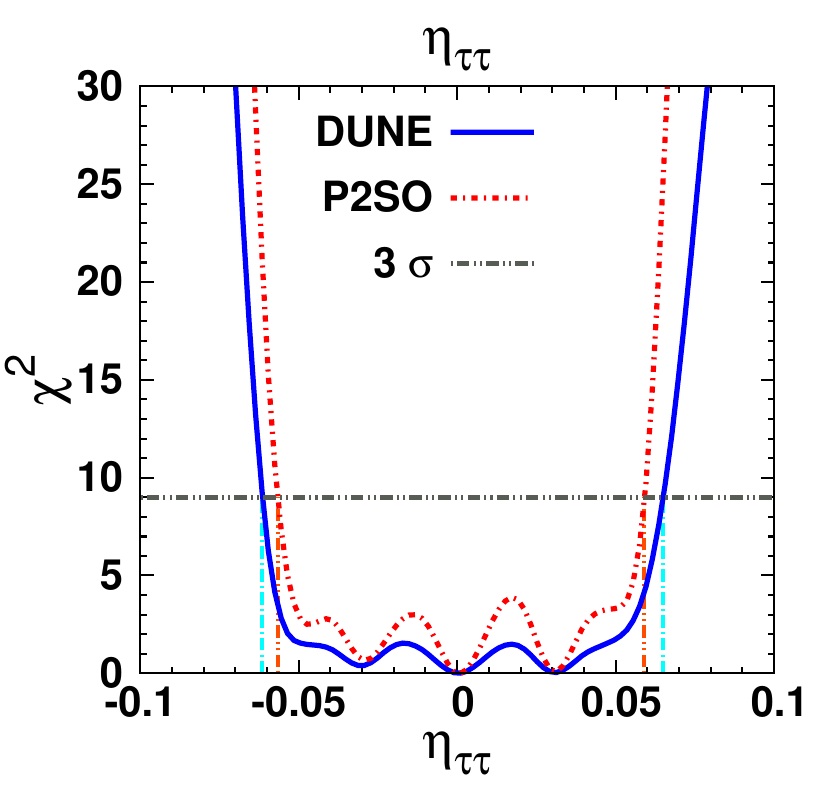}
    \caption{Bounds on the SNSI diagonal parameters ($\eta_{ee}, \eta_{\mu\mu}$ and $\eta_{\tau\tau}$) from DUNE and P2SO experiments. }
    \label{fig:bounds}
\end{figure}

In Fig. \ref{fig:bounds}, we have shown the bounds on the diagonal SNSI parameters adapting the methodology that we discussed in the previous paragraph. Left, middle and right panels show the bounds on $\eta_{ee}, \eta_{\mu\mu}$ and $\eta_{\tau\tau}$, respectively. In each panel, the blue curve represents the bound from DUNE while the red curve is for the bound from the P2SO experiment. 
From the figure we see that stringent bounds can be obtained on these parameters from both experiments. DUNE is better than P2SO to constraint the parameter $\eta_{ee}$ at $3\sigma$ C.L. While in the cases of $\eta_{\mu\mu}$ and $\eta_{\tau\tau}$, bounds from P2SO and DUNE are similar. The sensitivity limits on the parameters at $3\sigma$ C.L. are shown in Table \ref{tab:bound-table}. From the table, one can see that the bounds from DUNE and P2SO are more or less similar except the case of $\eta_{ee}$. 

\begin{table}
    \centering
    \begin{tabular}{|c|c|c|}
    \hline
       Parameters  & DUNE & P2SO  \\ \hline
        $\eta_{ee}$ & [-0.051, 0.048] & [-0.075, 0.073] \\ \hline
        $\eta_{\mu\mu}$ &[-0.0457,0.041] &[-0.042, 0.038] \\ \hline
        $\eta_{\tau\tau}$ & [-0.061, 0.065]& [-0.057, 0.059] \\ \hline    \end{tabular}
    \caption{Sensitivity limits on the SNSI parameters at $3\sigma$ C.L. from DUNE and P2SO experiment.}
   \label{tab:bound-table}
\end{table}
 
\begin{figure}
    \centering
    \includegraphics[scale=0.9]{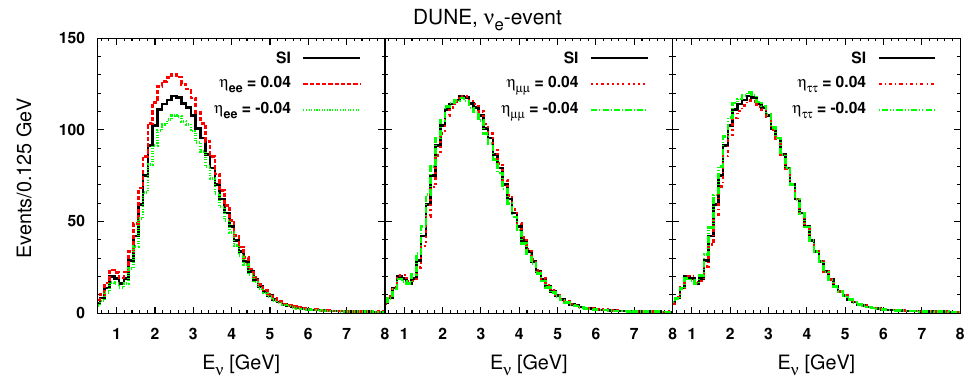}
    \includegraphics[scale=0.9]{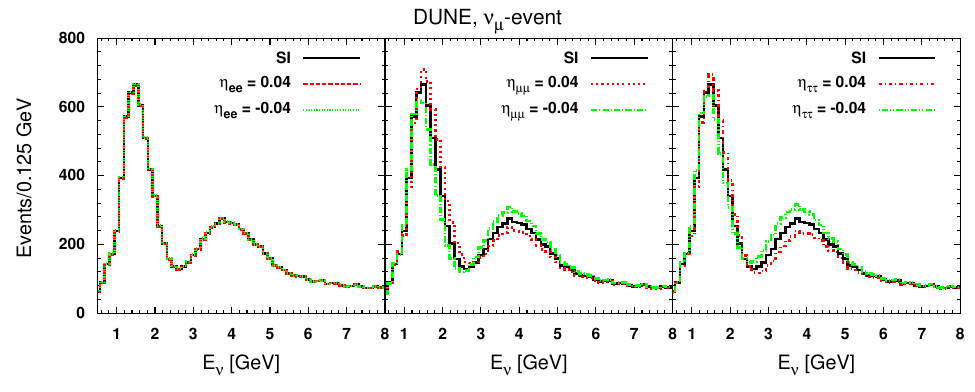}
    \caption{Number of neutrino events for appearance(disappearance) channel 
 as a function of neutrino energy in upper (lower) panel for DUNE experiment.}
    \label{fig:eventrates}
\end{figure}

To understand the contribution from the appearance channel and the disappearance channel in constraining the diagonal SNSI parameters, in Fig.~\ref{fig:eventrates} we show the neutrino events as a function of neutrino energy in standard case and in presence of SNSI for DUNE. For the calculation of events, we have used a value of $0.04$ for for all the three SNSI parameters i.e., $|\eta_{ee}|, |\eta_{\mu\mu}|$ and $|\eta_{\tau\tau}|$. This value of the SNSI parameter lies well within the $3 \sigma$ bound of these parameters. Upper (lower) panel is for the appearance (disappearance) events. Black-solid curves show the event-rates in standard case. Red and green dashed curves are the event-rates for $0.04$ and $-0.04$ values of SNSI parameters, respectively. 
From the figure, we realize that SNSI parameter $\eta_{ee}$ affects significantly to $\nu_e$ appearance events and mildly affects to  disappearance events. While in cases of $\eta_{\mu\mu}$ and $\eta_{\tau\tau}$ significant effect is observed for disappearance channel. Interestingly, the SNSI parameters  either enhance or deteriorate the event-rates depending upon the sign of the parameters. For example, the event-rates increased (decreased) for $+$ve ($-$ve) values of $\eta_{ee}$ in appearance channel. 

\begin{figure}
    \centering
    \includegraphics[scale=0.355]{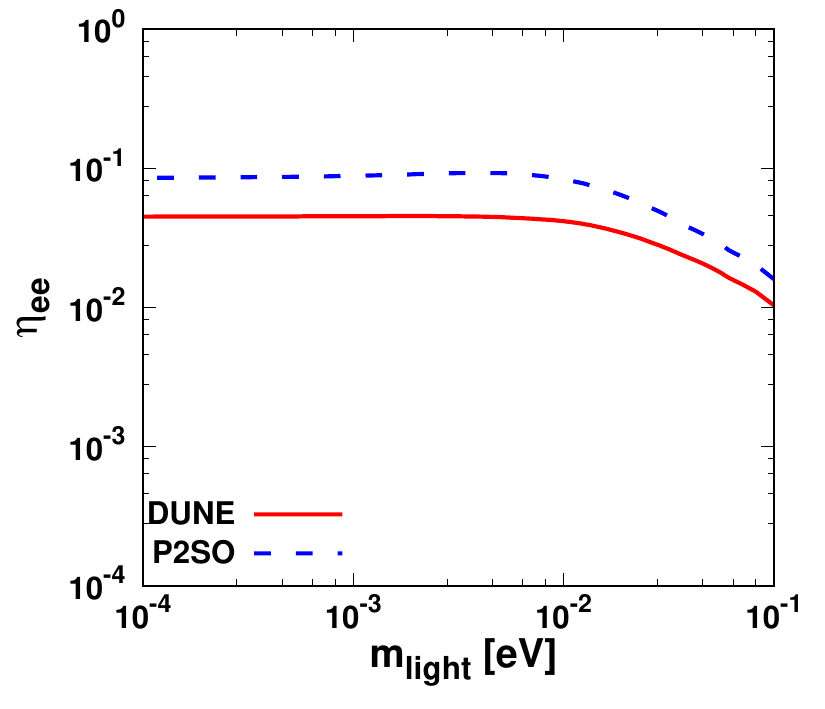}
    \includegraphics[scale=0.355]{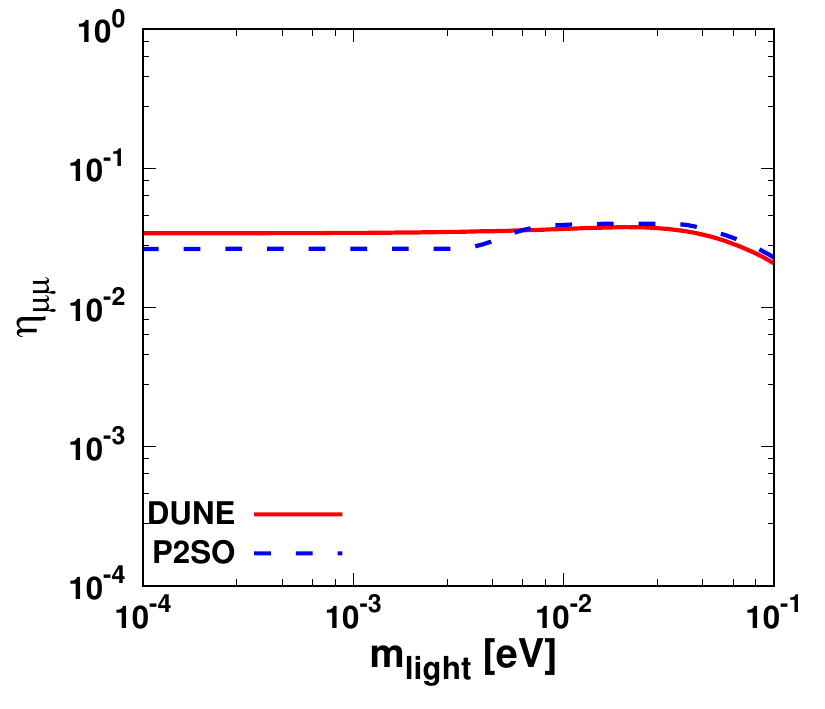}
    \includegraphics[scale=0.355]{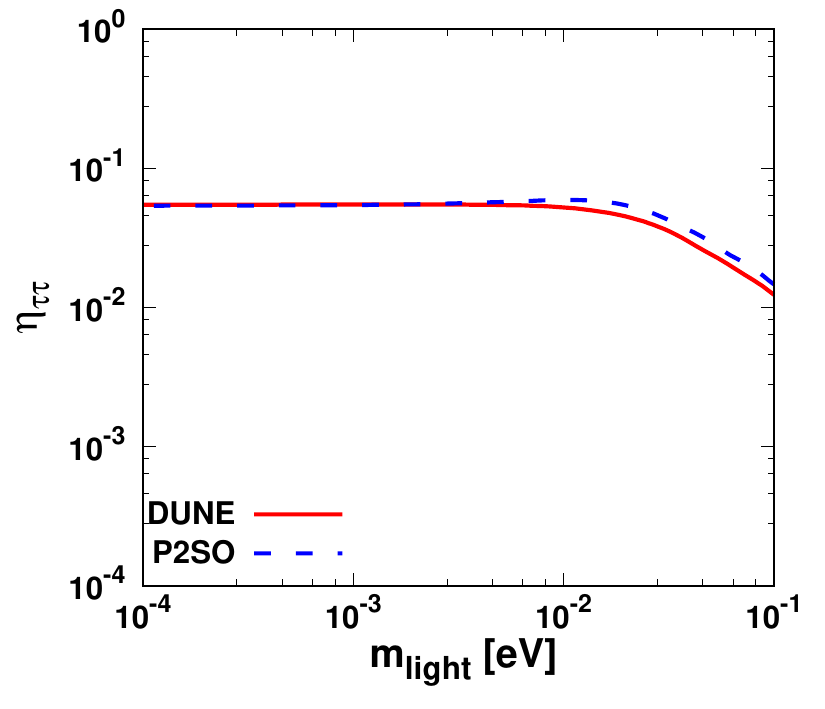}
    \caption{Constraints on scalar NSI parameters, $\eta_{ee}$ (left), $\eta_{\mu \mu}$ (middle) and $\eta_{\tau \tau}$ (right) for DUNE and P2SO experiments in normal mass ordering.}
    \label{fig:m1_bound}
\end{figure}

To see how the bounds on the diagonal SNSI parameters change with respect to the lowest neutrino mass $m_1$, in Fig.~\ref{fig:m1_bound} we have plotted the upper bound of the SNSI parameters at 3$\sigma$ as a function of $m_1$. The solid (dashed) lines represent the sensitivity of DUNE (P2SO) experiment.
We can notice from the figure that for the lightest neutrino mass below the order $10^{-2}$ eV, the constraints on SNSI parameters are almost unchanged. After that, the sensitivity of both DUNE and P2SO gets deteriorated. 

At this point, let us briefly discuss how these values of $m_1$ are compatible with the latest bound on the absolute neutrino masses. The most reliable bounds on the absolute neutrino masses come from cosmology which provides the upper bound on the sum of neutrino masses. The most relaxed bound on the sum of neutrino masses obtained from the Planck data, given as  $\sum m_\nu \le 0.54$ eV at 95\% C.L. \cite{Planck:2018vyg}. According to this, the neutrino masses are allowed to be in the quasi-degenerate regime i.e., $m_1 \sim m_2 \sim m_3 \sim 0.1$ eV. This implies the range of $x$-axis which is shown in Fig.~\ref{fig:m1_bound} is consistent with the latest bound on the absolute neutrino masses.

\subsection{Effect of SNSI in the standard scenario}
\label{eff}

\begin{figure}[htb!]
    \centering
\includegraphics[scale=0.35]{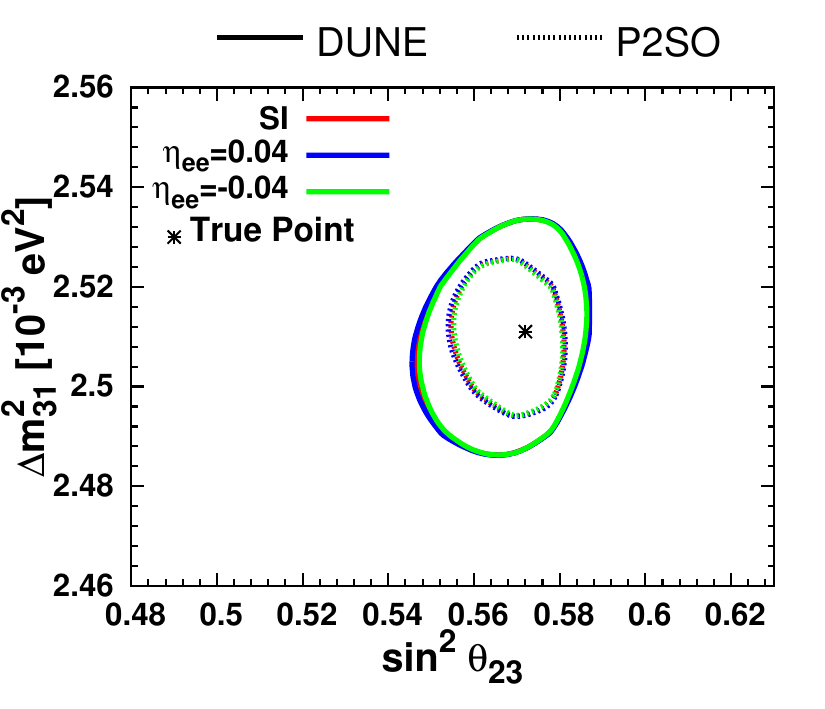}
\includegraphics[scale=0.35]{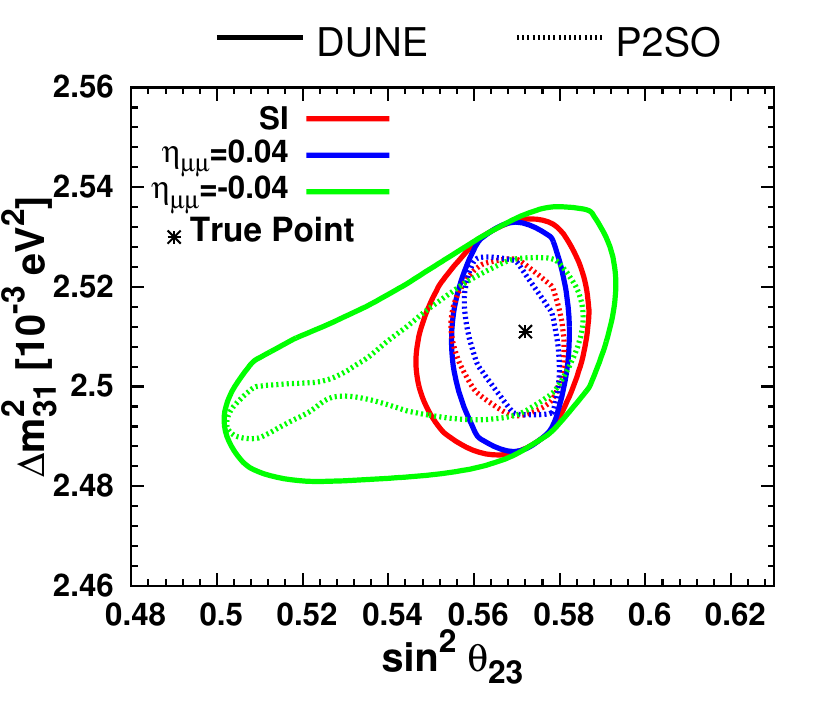}
\includegraphics[scale=0.35]{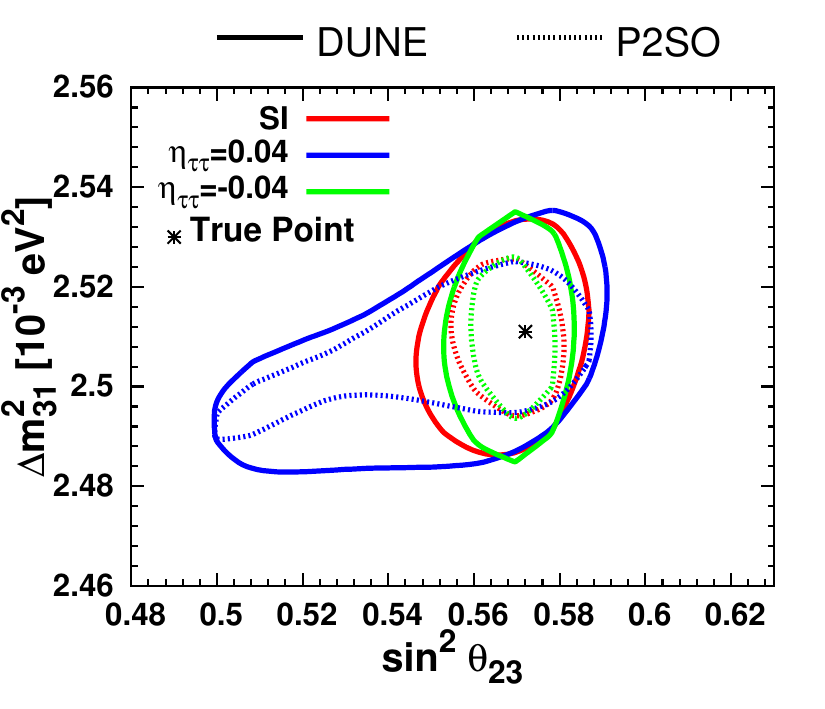}
    \caption{Allowed region between $\sin^2 \theta_{23} - \Delta m_{31}^2$ at $3\sigma$ C.L. in standard and in presence of SNSI parameters for DUNE and P2SO experiment. }
    \label{fig:allowed-region}
\end{figure}

Next we will study the effect of SNSI in the measurement of unknowns related to the standard oscillation parameters i.e., ordering of the neutrino masses, octant of the atmospheric mixing angle $\theta_{23}$ and the CP violation. This is usually done by taking the SNSI parameter in both true and test. Before doing that, first we will study the effect of SNSI in the precision measurement of $\theta_{23}$ and $\Delta m^2_{31}$. Fig.~\ref{fig:allowed-region} shows the allowed region $\sin^2\theta_{23}$ and $\Delta m^2_{31}$ in test parameter plane at $3\sigma$ C.L. by taking the value of all the three SNSI parameters to be $\pm 0.04$ in both true and test spectrum of the $\chi^2$. The true values of $\sin^2\theta_{23}$ and $\Delta m^2_{31}$ are denoted by a star. All the other parameters (except $\delta_{\rm CP}$) which are not shown in the panels are minimized by random sampling using their $1 \sigma$ error as a prior. The parameter $\delta_{\rm CP}$ is varied with a flat prior. Solid curves are for the DUNE experiment while dashed curves are for P2SO experiment. In each plot, the red colour curve represent the standard interaction case. Positive (negative) value of $\eta_{\alpha\beta}$ is represent by the blue (green) curve.   
In standard case, the allowed region is more constrained for P2SO experiment as compared to DUNE experiment. The impact of the three SNSI parameters are quite discernible from one another. In case of $\eta_{ee}$, there is no significant effect on the precision of $\theta_{23}$ and $\Delta m^2_{31}$. Interesting behaviour is observed for $\eta_{\mu\mu}$ and $\eta_{\tau\tau}$. For negative value of $\eta_{\mu\mu}$, the allowed parameter space increases whereas in case of $\eta_{\tau\tau}$, the allowed region increases for positive value of $\eta_{\tau\tau}$. However, in both the cases the sensitivity towards $\sin^2 \theta_{23}$ gets deteriorated but the precision of $\Delta m^2 _{31}$ does not change much. But most importantly we see that in this case the values of $\Delta m^2_{31}$ at $3 \sigma$ C.L. lie well within its current allowed ranges according to Nufit 5.2. Therefore, when one considers SNSI in both true and test, it is sufficient to minimize this parameter randomly with its $1 \sigma$ prior.  However, to be consistent with Fig.~\ref{fig:bounds}, in the following paragraphs, while studying the capability of P2SO and DUNE to measure the unknowns related to the standard oscillation parameters in presence of SNSI, we will minimize $\Delta m^2_{31}$ systematically within its current $3 \sigma$ range. The SNSI parameters will be fixed at $\pm 0.04$ in both true and test.

\begin{figure}[htb!]
    \centering
\includegraphics[scale=0.35]{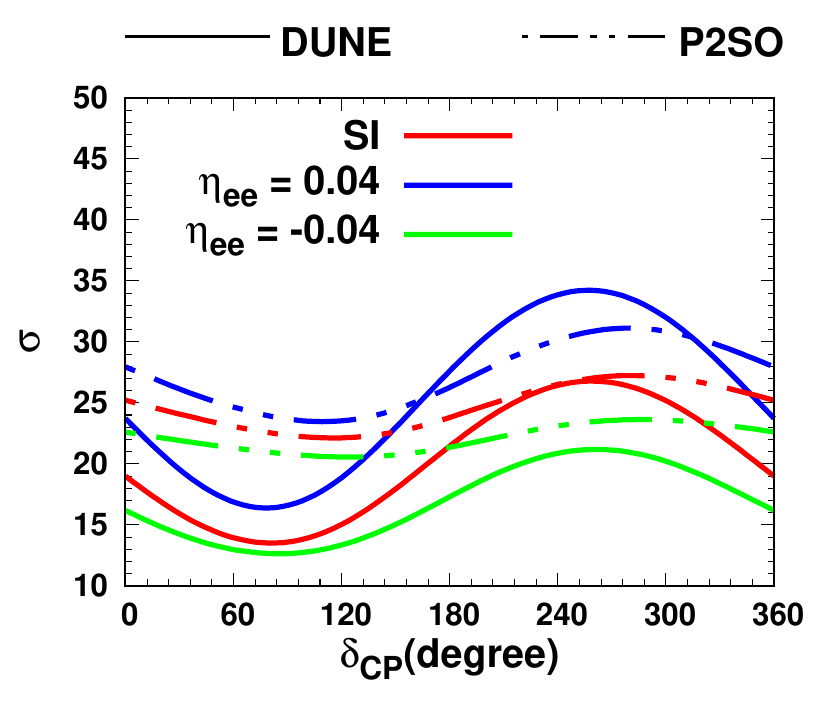}
\includegraphics[scale=0.35]{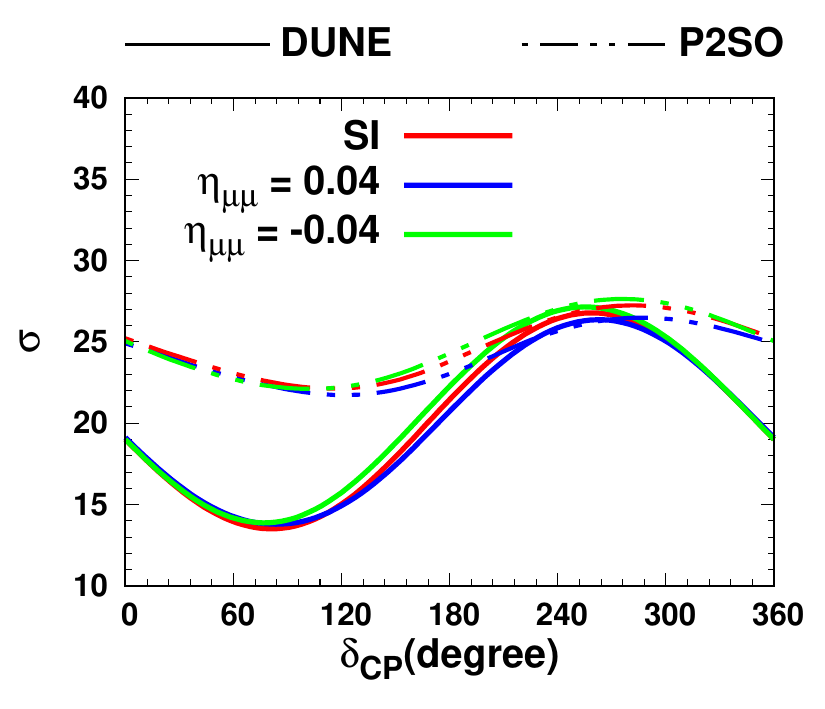}
\includegraphics[scale=0.35]{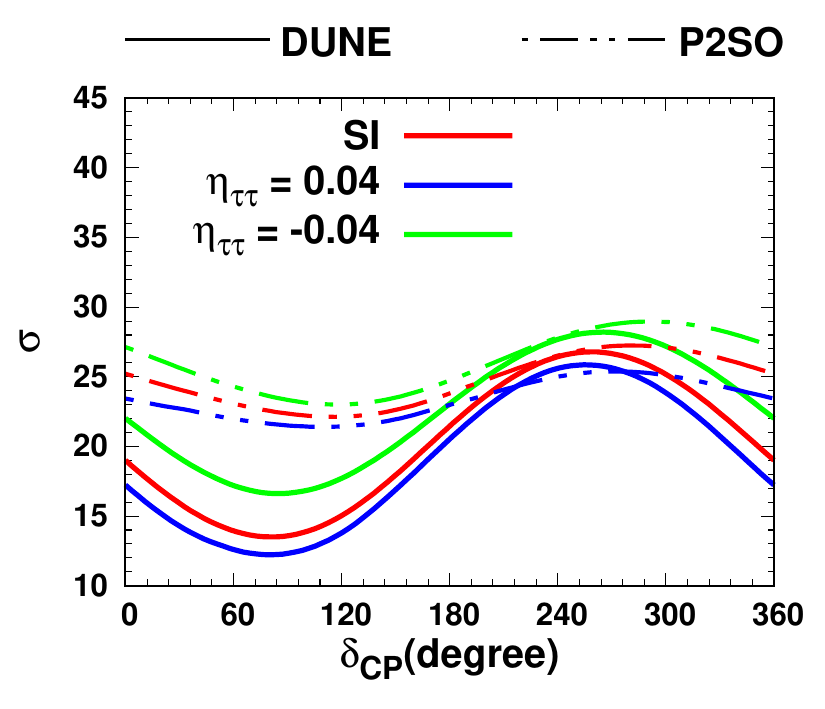}
    \caption{Mass hierarchy sensitivity of the SNSI diagonal parameters ($\eta_{ee}, \eta_{\mu\mu}$ and $\eta_{\tau\tau}$) for DUNE and P2SO experiment. }
    \label{fig:hierarchy-sensitivity}
\end{figure}

Let us first discuss the sensitivity to neutrino mass ordering in presence of SNSI which is shown in Fig.~\ref{fig:hierarchy-sensitivity} as function of true $\delta_{\rm CP}$. This figure shows the capability to determine the true nature of the neutrino mass ordering. This sensitivity is estimated by taking the normal mass ordering in true and the inverted mass ordering in the test. For inverted ordering we have taken $m_3 = 10^{-5}$ eV, $m_1 = \sqrt{m_3^2+\Delta m^2_{31}}$ and $m_2 = \sqrt{m_1^2+ \Delta m^2_{21}}$. Solid and dashed curves are for DUNE and P2SO, respectively. In each panel, standard case is represented by the red curve and blue (green) curve represents the $\eta_{\alpha\alpha}$ value as $+0.04~ (-0.04)$. Left, middle and right panels are for the parameters $\eta_{ee}$, $\eta_{\mu\mu}$ and $\eta_{\tau\tau}$, respectively.
From the figure, it is clear that both experiments have very high mass ordering sensitivity. For almost all the $\delta_{\rm CP}$ region, P2SO experiment has higher sensitivities than DUNE in the standard scenario.
In presence of SNSI, mass ordering sensitivity is affected more for $\eta_{ee}$ as compared to $\eta_{\mu\mu}$ and $\eta_{\tau\tau}$. After including SNSI parameters to standard case, the sensitivities may increase or decrease depending upon the sign of $\eta_{\alpha\beta}$. The sensitivity is increased (decreased) as compared to the standard three flavour scenario for a positive (negative) value of $\eta_{ee}$. However, it is opposite for $\eta_{\tau\tau}$. This can be better understood from Tab.~\ref{table_events_sg}.

In Table~\ref{table_events_sg}, we have listed the difference of event numbers ($N$) between the normal ordering and inverted ordering coming from the appearance channel corresponding to $\delta_{\rm CP}$ (true)=$90^\circ$ for both P2SO and DUNE. The events for normal ordering is calculated at the true values of the oscillation parameters and the events for inverted ordering are calculated corresponding to the $\chi^2$ minimum.  
\begin{table}
\centering
\begin{tabular}{ |p{3cm}|p{3cm}|p{3cm}|p{3cm}|  }
\hline
\multicolumn{4}{|c|}{Appearance channel event difference (Normal ordering - Inverted ordering)} \\
\hline
Experiments& $\eta = 0.04$ & $\eta = 0$ & $\eta = - 0.04$ \\
\hline
P2SO ($\eta_{ee}$)& 6992 & 6273 & 5609 \\
DUNE ($\eta_{ee}$)& 360 & 244 & 150 \\
\hline
P2SO ($\eta_{\tau \tau}$)& 5336 & 6273 & 7510 \\
DUNE ($\eta_{\tau \tau}$)& 151 & 244 & 343 \\
\hline
\end{tabular}
\caption{Appearance channel event difference for P2SO and DUNE for $\delta^{\rm true}_{CP} = 90^{o}$. These events corresponds to 3 years running of P2O and 6.5 year running of DUNE. }
\label{table_events_sg}
\end{table}
From this table, we see that $N$ is higher (lower) when $\eta_{ee}$ is positive (negative) as compared to the standard case but this is opposite for $\eta_{\tau\tau}$. This explains why mass ordering sensitivity is higher (lower) for a positive (negative) value of $\eta_{ee}$ and a negative (positive) value of $\eta_{\tau\tau}$.

Next let us discuss the CP violation (CPV) discovery sensitivity\footnote{Note that though here we only discuss the CPV for the diagonal SNSI parameters, the contribution for the CPV will mostly arise from the off-diagonal SNSI parameters which are complex in nature. We have included a section in the appendix to discuss this.}. Fig.~\ref{fig:CPV-sensitivity} shows the CPV sensitivities as a function of true $\delta_{\rm CP}$ for DUNE and P2SO experiments. In particular, this figure shows the ability of the experiments to exclude the CP conserving values of $\delta_{\rm CP}$. For each true $\delta_{\rm CP}$, we have obtained the minimum $\chi^2$ for test $\delta_{\rm CP}$ as CP conserving values ($0^\circ$ and $180^\circ$). Solid and dashed curves show the sensitivities of DUNE and P2SO, respectively. In each plot, the red curves are for standard interaction cases, while the blue and green curves are for SNSI parameters as $+0.04$ and $-0.04$, respectively. The left, middle and right panels are for the SNSI parameter $\eta_{ee}$, $\eta_{\mu\mu}$ and $\eta_{\tau\tau}$. 
 \begin{figure}[htb!]
    \centering
\includegraphics[scale=0.35]{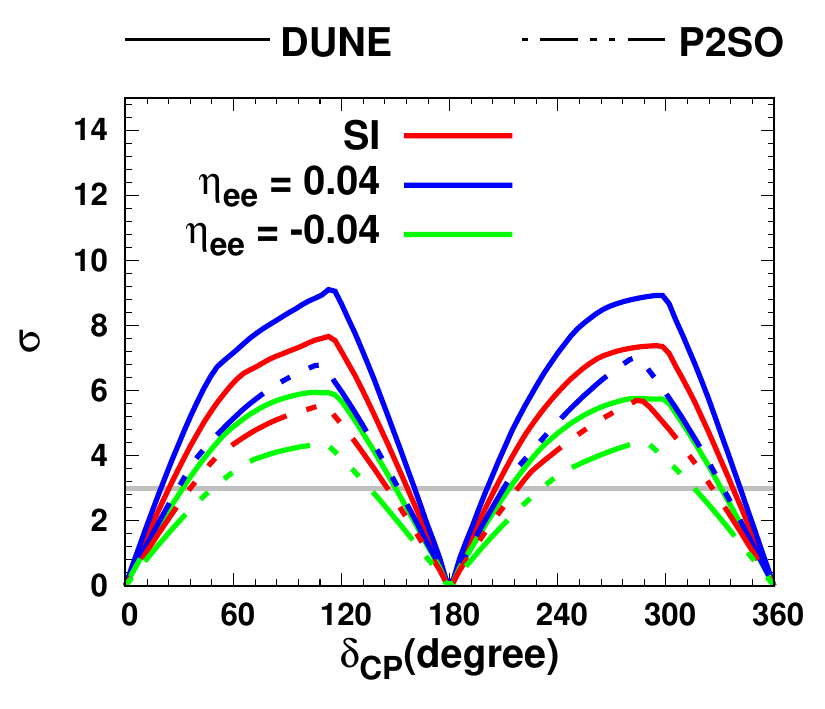}
\includegraphics[scale=0.35]{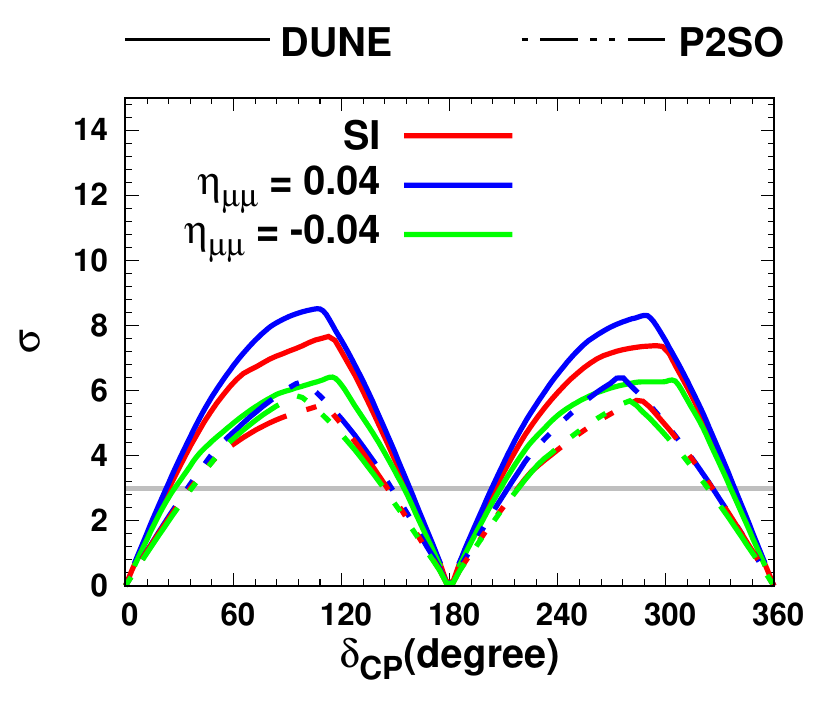}
\includegraphics[scale=0.35]{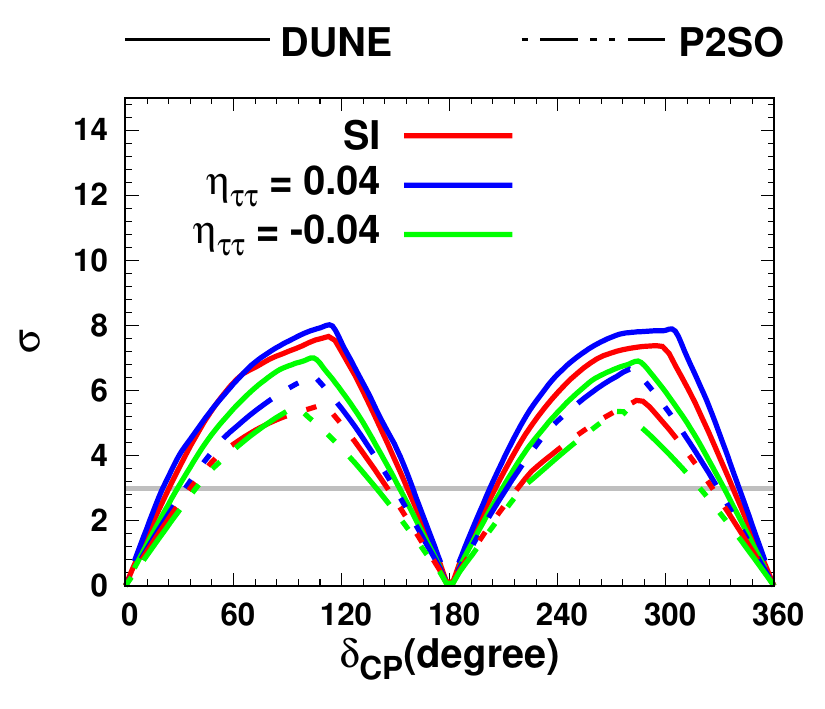}
    \caption{CPV sensitivity of the SNSI diagonal parameters ($\eta_{ee}, \eta_{\mu\mu}$ and $\eta_{\tau\tau}$) for DUNE and P2SO experiment. }
    \label{fig:CPV-sensitivity}
\end{figure}
From the figures one can conclude that in standard case the DUNE experiment will have higher CPV sensitivities compared to P2SO experiment. The effect of SNSI parameters is different for each parameter. The parameter $\eta_{ee}$ and $\eta_{\mu\mu}$ have significantly higher effect on the CPV sensitivity than $\eta_{\tau\tau}$. Positive value of $\eta_{\alpha\alpha}$ increases the CPV sensitivities whereas the sensitivities get decreased for negative value of these parameters. 

\begin{figure}[htb!]
    \centering
\includegraphics[scale=0.35]{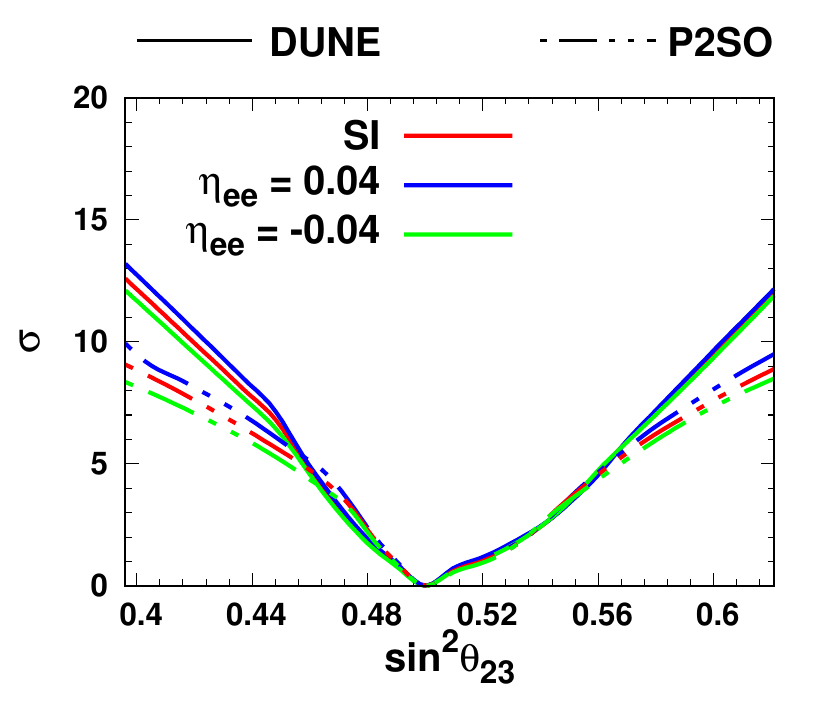}
\includegraphics[scale=0.35]{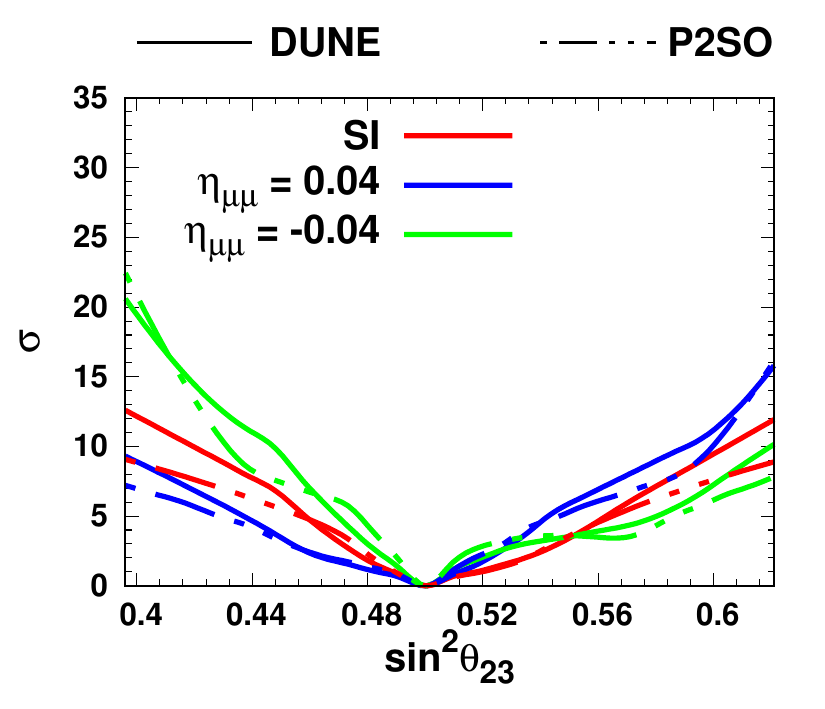}
\includegraphics[scale=0.35]{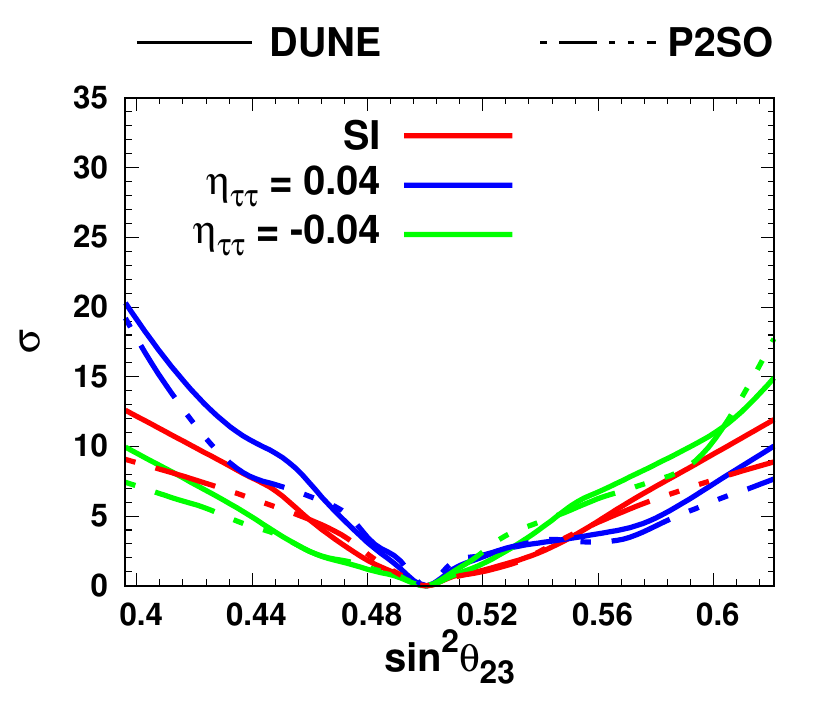}
    \caption{Octant sensitivity of the SNSI diagonal parameters ($\eta_{ee}, \eta_{\mu\mu}$ and $\eta_{\tau\tau}$) for DUNE and P2SO experiments. }
    \label{fig:Octant-sensitivity-1}
\end{figure}

Finally, let us discuss the octant sensitivity of $\theta_{23}$ in presence of SNSI. Octant sensitivity is the capability to exclude the degeneracy between lower octant (LO) i.e., $\theta_{23} < 45^\circ$ and higher octant (HO) i.e., $\theta_{23} > 45^\circ$. 
Fig. \ref{fig:Octant-sensitivity-1} shows the octant sensitivity as a function of true $\sin^2\theta_{23}$.  This figure is generated in the following way. If the true $\sin^2\theta_{23}$ is in LO (HO), then test $\sin^2\theta_{23}$ varied in HO (LO). In each panel, standard case is represented by the red curve  and blue (green) curve represents the $\eta_{\alpha\alpha}$ as $+0.04~ (-0.04)$. Left, middle and right panels are for the parameters $\eta_{ee}$, $\eta_{\mu\mu}$ and $\eta_{\tau\tau}$, respectively. 
One can conclude form the figure that, in the standard case  the sensitivity of DUNE is slightly higher than P2SO. Considering the SNSI parameters, the parameter $\eta_{ee}$ affects the least to the sensitivity. Parameter $\eta_{\mu\mu}$ and $\eta_{\tau\tau}$ have significant effect on the sensitivity. In case of $\eta_{\mu\mu}$, the sensitivity is enhanced for the positive value and deteriorated for the negative value. This observation is reversed in presence of parameter $\eta_{\tau\tau}$.

\section{Conclusion}

In this paper, we have studied the non-standard interaction mediated by a scalar field (SNSI) in the context of the P2SO experiment and compared its sensitivity with the DUNE experiment. P2SO is a proposed long-baseline neutrino experiment at the KM3NeT facility whereas DUNE is the upcoming long-baseline experiment at Fermilab. In the presence of SNSI, the neutrino masses in Hamiltonian of the neutrino oscillation gets a correction term which in turn alters the probabilities of the neutrino oscillation. Interestingly, in this case, the neutrino oscillation probabilities become function of the absolute neutrino masses. Due to the introduction of SNSI, the neutrino oscillation parameter space is increased by 3 real SNSI parameters and 3 complex parameters. In this study our aim is to study the capability of P2SO and DUNE to constrain the real SNSI parameters i.e., $\eta_{ee}$, $\eta_{\mu\mu}$ and $\eta_{\tau\tau}$ and also study how these parameters impact the measurement of the standard oscillation parameters by these two experiments. We have presented our results considering the normal ordering of the neutrino masses.

At the probability level, we have shown that the sensitivity of $\eta_{ee}$ is expected to mainly come from the appearance channel whereas the sensitivities to $\eta_{\mu\mu}$ and $\eta_{\tau\tau}$ are expected to come from the disappearance channel. However, some sensitivity to $\eta_{\mu\mu}$ and $\eta_{\tau\tau}$ can also come from the appearance channel. While estimating the capability of DUNE and P2SO to constrain the SNSI parameters, we find that the parameter $\Delta m^2_{31}$ has a non-trivial role for $\eta_{\mu\mu}$ and $\eta_{\tau\tau}$ when one considers standard three flavour scenario in the true spectrum of the $\chi^2$ and SNSI in the test spectrum of the $\chi^2$. For these two parameters, the $\chi^2$ minimum can appear outside the current allowed values of $\Delta m^2_{31}$. Therefore, the bounds on these two SNSI parameters depend upon how $\Delta m^2_{31}$ is minimized in the analysis. In our work, we choose to minimize this parameter systematically within its current three sigma range. In our analysis, we find that for $\eta_{ee}$, the bound from DUNE is stronger as compared to the bound from P2SO.  For the other two parameters, the bounds from DUNE and P2SO are comparable. From the event level analysis, we found that for $\eta_{ee}$, the events from appearance channel mostly contribute in the sensitivity whereas for $\eta_{\mu\mu}$ and $\eta_{\tau\tau}$, the events from disappearance channel mostly contribute in the sensitivity. Further, we also find that if we vary the lowest neutrino mass i.e., $m_1$, then for $m_1 < 0.01$ eV, the bounds of the SNSI parameters do not change. However, when one increases $m_1$ beyond 0.01 eV, the sensitivity starts to fall. 

While estimating the effect of the SNSI parameters on the measurement of standard oscillation parameters, we find that $\chi^2$ minimum always comes within the current allowed values of $\Delta m^2_{31}$ when one considers SNSI in both true and test spectrum of the $\chi^2$. The value of the three SNSI parameters that we consider in our analysis is $\pm 0.04$. Additionally, we find that in the presence of SNSI, the precision of $\theta_{23}$ gets deteriorated significantly for $\eta_{\mu\mu}$ and $\eta_{\tau\tau}$ but the precision of $\Delta m^2_{31}$ does not get much affected. For $\eta_{ee}$, the precision of  $\Delta m^2_{31}$ and $\theta_{23}$ is same as in the case of standard three flavour scenario. Going ahead, the change in the neutrino mass ordering sensitivity and CP violation sensitivity due to SNSI is maximum for $\eta_{ee}$ as compared to $\eta_{\mu\mu}$ and $\eta_{\tau\tau}$ whereas the change in the  octant sensitivity is maximum for $\eta_{\tau\tau}$. These sensitivities can be either higher or lower than the standard three flavour scenario depending on the relative sign of the SNSI parameters.

\acknowledgments
DKS would like to acknowledge Prime Minister's Research Fellowship, Govt. of India. MG would like to thank Alessio Giarnetti for discussions regarding scalar NSI. This work has been in part funded by Ministry of Science and Education of Republic of Croatia grant No. KK.01.1.1.01.0001. RM would like to acknowledge University of Hyderabad IoE project grant no. RC1-20-012. We gratefully acknowledge the use of CMSD HPC facility of University of Hyderabad to carry out the computational works. LP would like to acknowledge DAE and DST, Govt. of India, for research fellowship. We thank Sambit K. Pusty, Shivaramakrishna Singirala and Aman Gupta for useful discussions. 

\appendix
\section{Separation of SNSI from VNSI}

\begin{figure} [htb!]
    \centering
    \includegraphics[scale=0.9]{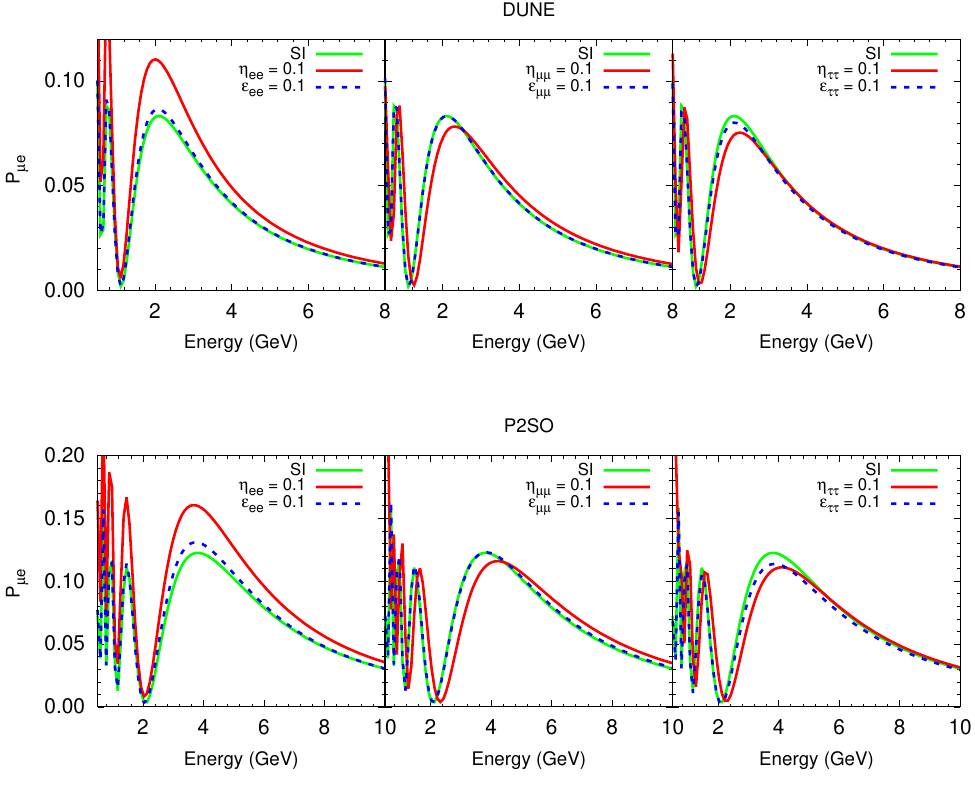}
    \caption{Aappearance probability as a function of neutrino energy for DUNE and P2SO experiment in presence of diagonal SNSI and VNSI parameters.}
    \label{s_vs_v}
\end{figure}

As mentioned in section \ref{bg}, the neutrino oscillation Hamiltonian in presence of SNSI parameter $\eta$ can be written as:

\begin{eqnarray}
  H_{SNSI} = E_\nu + \frac{\mathcal{M} \mathcal{M}^\dagger}{2E_\nu} + {\rm diag}(\sqrt{2}G_F N_e,0,0) \;,
  \label{ham_s}
\end{eqnarray}
with
\begin{eqnarray}
 \mathcal{M} = U~{\rm diag}(m_1, m_2, m_3)~U^\dagger +  \sqrt{|\Delta m^2_{31}|}
\begin{pmatrix}
\eta_{ee} & \eta_{e\mu} & \eta_{e\tau}\\
\eta_{\mu e} & \eta_{\mu \mu} & \eta_{\mu \tau} \\
\eta_{\tau e} & \eta_{\tau \mu} & \eta_{\tau\tau}
\end{pmatrix}\;.
 \end{eqnarray}
Whereas the same Hamiltonian in presence of VNSI parameters $\epsilon$ can is written as:
\begin{eqnarray}
  H_{VNSI} = E_\nu + \frac{1}{2E_\nu} U~{\rm diag}(m_1, m_2, m_3)~U^\dagger+\sqrt{2}G_F N_e\begin{pmatrix}
1+\epsilon_{ee} & \epsilon_{e\mu} & \epsilon_{e\tau}\\
\epsilon_{\mu e} & \epsilon_{\mu \mu} & \epsilon_{\mu \tau} \\
\epsilon_{\tau e} & \epsilon_{\tau \mu} & \epsilon_{\tau\tau}
\end{pmatrix}\; \;.
  \label{ham_v}
\end{eqnarray}
 Comparing Eqs.\ref{ham_s} and \ref{ham_v}, we see that SNSI has been added as a correction term in the mass matrix whereas VSNSI is added as a correction term in the matter potential. To understand their effect in the neutrino oscillation probabilities, in Fig.~\ref{s_vs_v}, we have shown the  appearance channel probability as a function neutrino energy for both DUNE (top row) and P2SO (bottom row) in presence of diagonal SNSI (red curve) and VNSI (blue curve) parameters. These panels clearly show that for same values of the SNSI and VNSI parameters, the values of their corresponding probabilities are very different.

\section{Appearance channel probability expression for $\eta_{ee}$}

To derive the probability formula in presence of SNSI for $\eta_{ee}$, we have followed the formalism as mentioned in Ref.~\cite{Ioannisian:2018qwl}. The effective Hamiltonian in the presence of SNSI is mentioned in Eq. \ref{ham} and can be written in expanded form as 
\begin{eqnarray}
    H_{SNSI} &\approx& \frac{1}{2 E_{\nu}}\left[(M + \delta M)(M + \delta M)^{\dagger} + 2 E_{\nu} V \right] \nonumber \\
    &=& \frac{1}{2 E_{\nu}}\left[M M^{\dagger} + \delta M \delta M^{\dagger} + M \delta M^{\dagger} + \delta M M^{\dagger} + 2 E_{\nu} V \right],
\end{eqnarray}

where
\begin{eqnarray}
   M &=& U
\begin{pmatrix}
m_1 & 0 & 0\\
0 & m_2 & 0 \\
0 & 0 & m_3
\label{eta}
\end{pmatrix}\;U^{\dagger}, ~
\delta M =  \sqrt{|\Delta m^2_{31}|}
\begin{pmatrix}
\eta_{ee} & 0 & 0\\
0 & 0 & 0 \\
0 & 0 & 0
\label{eta}
\end{pmatrix}\;, \nonumber \\~
V &=& V_m 
\begin{pmatrix}
1 & 0 & 0\\
0 & 0 & 0 \\
0 & 0 & 0
\label{eta}
\end{pmatrix}\;,
\end{eqnarray}
with $U$ = $R_{23} U^{\dagger}_{\delta} R_{13} U_{\delta} R_{12}$ and $V_m$ = $\sqrt{2} G_{f} N_{e}$. Here $R_{ij}$ is the rotation matrix in (i,j) plane and $U_{\delta}$ = ${\rm diag}\left(1,1,e^{i \delta_{CP}}\right)$. We have used normal mass ordering for our analysis, so for the matter of convenience we will use $\sqrt{|\Delta m^2_{31}|}$ = $\sqrt{\Delta m^2_{31}}$ for further calculations. The time evolution of the neutrinos in their flavor state is given by the equation

\begin{eqnarray}
    i \frac{\partial }{\partial t} \ket{\nu_{\alpha}} &=& H_{SNSI} \ket{\nu_{\alpha}},
\end{eqnarray}
and the probability amplitude of their oscillations is given by the $S$ matrix
\begin{eqnarray}
    S &=& T e^{- i \int_{x_i}^{x_f} H_{SNSI} (x) \,dx},
\end{eqnarray}
where the $T$ represents the space ordering. Considering constant matter density, it is convenient to write the $S$ matrix element as 
\begin{eqnarray}
    S_{\alpha \beta} &=& \left( e^{- i U_{\rm eff} H^{d}_{SNSI} U^{\dagger}_{\rm eff} (x_f - x_i)} \right)_{\alpha \beta} 
    = \left( U_{\rm eff}\hspace*{0.08 true cm} e^{-i H^{d}_{SNSI} L} \hspace*{0.08 true cm}U^{\dagger}_{\rm eff} \right)_{\alpha \beta}.
\end{eqnarray}
Here $L$ is the neutrino travel distance via the earth matter, $U_{\rm eff}$ is the effective $PMNS$ matrix in the presence of SNSI in the matter and $H^{d}_{SNSI}$ is the diagonalised effective Hamiltonian of the form 
\begin{eqnarray}
    H^{d}_{SNSI} &=&
    \begin{pmatrix}
H_1 & 0 & 0\\
0 & H_2 & 0 \\
0 & 0 & H_3
\label{effham}
\end{pmatrix}\;.
\end{eqnarray}
Defining $\Delta^{\rm eff}_{21}$ = $H_2 - H_1$ and $\Delta^{\rm eff}_{31}$ = $H_3 - H_1$, we can rewrite the $S$ matrix element by ignoring the common phase $e^{-iH_1 L}$ as
\begin{eqnarray}
    S_{\alpha \beta} &=& \left[ U_{\rm eff} 
    \begin{pmatrix}
1 & 0 & 0\\
0 & e^{-i \Delta^{\rm eff}_{21} L} & 0 \\
0 & 0 & e^{-i \Delta^{\rm eff}_{31} L}
\label{effham}
\end{pmatrix}\;
    U^{\dagger}_{\rm eff}\right]_{\alpha \beta}.
    \label{Sab}
\end{eqnarray}
For the simplicity of calculations, we can change the flavor basis to some auxiliary basis $\ket{\widetilde{\nu_{\alpha}}}$ = $U^{a} \ket{\nu_{\alpha}}$ by some auxiliary rotation $U^{a}$ = $R_{23} U^{\dagger}_{\delta} R_{13}$ which will change the effective Hamiltonian and $S$ matrix to 
\begin{eqnarray}
    H^{'}_{SNSI} &=& U^{a \dagger} \hspace*{0.08 true cm}H_{SNSI}\hspace*{0.08 true cm} U^{a} ~{\rm and}~ S = U^{a} \hspace*{0.08 true cm}e^{-i H^{'}_{SNSI} L} \hspace*{0.08 true cm}U^{a \dagger},
    \label{prm}
\end{eqnarray}
respectively.
This new effective Hamiltonian is now independent of $\delta_{CP}$ and can be easily diagonalized by two consecutive rotations in (1,3) and (1,2) plane 
\begin{eqnarray}
    R^{{\rm '} T}_{12} R^{{\rm '} T}_{13}\hspace*{0.08 true cm} H^{\rm '}_{SNSI}\hspace*{0.08 true cm} R^{\rm '}_{13} R^{\rm '}_{12} &=&
    \begin{pmatrix}
H_1 & 0 & 0\\
0 & H_2 & 0 \\
0 & 0 & H_3
\label{diagHam}
\end{pmatrix}\;.
\label{digH}
\end{eqnarray}
Here we have neglected the (2,3) and (3,2) elements which are generated after the (1,3) and (1,2) rotations. The additional rotation angles $\theta^{'}_{1 3}$ and $\theta^{'}_{1 2}$ for which the Eq. \ref{diagHam} holds are given in Eq. \ref{13} and Eq. \ref{12} respectively. Finally the effective mixing matrix in the presence of SNSI in matter can be expressed as
\begin{eqnarray}
    U_{\rm eff} &=& R_{23} U^{\dagger}_{\delta} R_{13} R^{\rm '}_{13} U_{\delta} R^{\rm '}_{12}
    \label{effU}
\end{eqnarray}
Substituting Eq. \ref{effU} and Eq. \ref{digH} in Eq. \ref{prm}, we will easily recover Eq. \ref{Sab}. As the $S$ matrix elements represent the probability amplitudes, the probability of transition from $\ket{\nu_{\alpha}}$ flavor to $\ket{\nu_{\beta}}$ flavor is expressed as $P_{\alpha \beta}$ = $|S_{\alpha \beta}|^2$. Using the above equations, the appearance channel probability expression for $\eta_{ee}$ becomes:
\begin{eqnarray}
    P_{\mu e} &=& \cos^2 \left(\theta_{13} + \theta^{'}_{13}\right) \cos^2 \theta_{23} \sin^2 2\theta^{'}_{12} \sin^2 \left(\frac{\Delta^{\rm eff}_{21} L}{2}\right) + \frac{1}{16} \sin^2 2\left(\theta_{13} + \theta^{'}_{13}\right) \sin^2 \theta_{23} \nonumber \\
    &\times&  \Bigg\{ \Bigg. 7 + \cos \left( \Delta^{\rm eff}_{21} L \right) - 4 \cos \left((\Delta^{\rm eff}_{21}-\Delta^{\rm eff}_{31}) L \right) - 4 \cos \left( \Delta^{\rm eff}_{31} L \right) + 2 \cos 4\theta^{\rm '}_{12} \sin^2 \left(\frac{\Delta^{\rm eff}_{21} L}{2}\right) \nonumber \\
    &-& 8 \cos 2\theta^{\rm '}_{12} \sin \left(\frac{\Delta^{\rm eff}_{21} L}{2}\right) \sin \left(\frac{(\Delta^{\rm eff}_{21}-2\Delta^{\rm eff}_{31})L}{2}\right) \Bigg. \Bigg\} + P^{\delta_{CP}}_{\mu e} 
    \label{prob}
\end{eqnarray}
where $P^{\delta_{CP}}_{\mu e}$ is the CP phase dependent part and is expressed as
\begin{eqnarray}
    \label{dcp}
    P^{\delta_{CP}}_{\mu e} &=& \cos^2 \left(\theta_{13} + \theta^{'}_{13}\right) \sin (2 \theta_{23}) \sin \left(\theta_{13} + \theta^{'}_{13}\right) \sin (2 \theta^{\rm '}_{12}) \sin \left(\frac{\Delta^{\rm eff}_{21} L}{2}\right) \\ \nonumber
    &\times& \left[ \cos \delta_{CP} \cos 2 \theta^{\rm '}_{12} \sin \left( \frac{\Delta^{\rm eff}_{21} L}{2} \right) -  \cos \left( \frac{\Delta^{\rm eff}_{21} L}{2} \right) \sin \delta_{CP} + \sin \left(  \delta_{CP} + \Delta^{\rm eff}_{31} L - \frac{\Delta^{\rm eff}_{21} L}{2} \right)  \right] \nonumber
\end{eqnarray}
In the above equations, the additional rotation in {1,3} plane can be expressed as
\begin{eqnarray}
    \sin 2 \theta^{'}_{13} = \frac{a_{13}}{\sqrt{a^{2}_{13} + b^{2}_{13}}}
    \label{13}
\end{eqnarray}
where 
\begin{eqnarray}
    a_{13} &=& \left[ 2 V_m + 2 \Delta m^{2}_{31} \eta^{2}_{ee} + (m_1 + m_2 + 2 m_3) \sqrt{\Delta m^{2}_{31}} \eta_{ee} + \sqrt{\Delta m^{2}_{31}} \eta_{ee} (m_1 - m_2) \cos 2 \theta_{12} \right] \sin 2 \theta_{13}, \nonumber \\
    b_{13} &=& 2 \bigg[ \bigg. \Delta m^{2}_{31} - \Delta m^{2}_{21} \sin^2 \theta_{12} - V_m \cos 2 \theta_{13} + 2 m_3 \sqrt{\Delta m^{2}_{31}} \eta_{ee} \sin^2 \theta_{13}  \nonumber \\
    &-& \Delta m^{2}_{31} \eta^{2}_{ee} \cos 2 \theta_{13} - 2 \sqrt{\Delta m^{2}_{31}} \eta_{ee} \cos^2 \theta_{13} \left( m_1 \cos^2 \theta_{12} +  m_2 \sin^2 \theta_{12} \right) \bigg. \bigg]. 
\end{eqnarray}
The additional rotation in the ${1,2}$ plane can be expressed as
\begin{eqnarray}
    \sin 2 \theta^{\rm '}_{12} &=& \frac{- a_{12}}{\sqrt{a^{2}_{12} + b^{2}_{12}}}\;, 
\label{12}
\end{eqnarray}
where,
\begin{eqnarray}
    a_{12}  &=& \sin 2 \theta_{12} \left[ 2 \Delta m^{2}_{21} \cos \theta^{'}_{13} + \sqrt{\Delta m^{2}_{31}} \eta_{ee} (m_2 - m_1) \left(\cos \theta^{'}_{13} + \cos (2 \theta_{13} + \theta^{'}_{13}) \right)  \right], \nonumber \\
    b_{12} &=& 2 \Bigg\{ \Bigg. - \cos^2 \theta_{12} \left( \Delta m^{2}_{21} - 2 \sqrt{\Delta m^{2}_{31}} m_1 \eta_{ee} \cos \theta_{13} \cos \theta^{'}_{13} \cos (\theta_{13} + \theta^{'}_{13}) \right) \nonumber \\
    &+& \cos^2 \theta^{'}_{13} \left( \Delta m^{2}_{21} \sin^2 \theta_{12} + \cos^2 \theta_{13} \left[ \Delta m^{2}_{31} \eta^{2}_{ee} + V_m + 2 \sqrt{\Delta m^{2}_{31}} \eta_{ee} m_2 \sin^2 \theta_{12} \right] \right) \nonumber \\
    &+& \sin^2 \theta^{'}_{13} \left(\Delta m^{2}_{31} + \sin^2 \theta_{13} \left[\Delta m^{2}_{31} \eta^{2}_{ee} + V_m + 2 \sqrt{\Delta m^{2}_{31}} \eta_{ee} m_3 \right] \right) \Bigg. \Bigg\} \nonumber \\
    &-& \frac{1}{2} \bigg[ \bigg. 2 \Delta m^{2}_{31} \eta^{2}_{ee} + \sqrt{\Delta m^{2}_{31}} \eta_{ee} (m_2 + 2 m_3) + 2 V_m \nonumber \\
    &-& \sqrt{\Delta m^{2}_{31}} \eta_{ee} m_2 \cos 2 \theta_{12} \bigg. \bigg] \sin 2 \theta_{13} \sin 2 \theta^{'}_{13}\;. 
\end{eqnarray}
The effective mass square splittings are 
\begin{eqnarray}
    \Delta^{\rm eff}_{21} 
    &=& \frac{1}{8 E} \Bigg[ \Bigg. \cos 2 \theta^{'}_{12} \Bigg\{ \Bigg. \Delta m^{2}_{21} - 2 \Delta m^{2}_{31} \left( 1 + \eta^{2}_{ee} \right) - \sqrt{\Delta m^{2}_{31}} \eta_{ee} \left( m_1 + m_2 + 2 m_3 \right) \nonumber \\
    &-& 2V_m + \left( 2 \Delta m^{2}_{31} - \Delta m^{2}_{21} \right) \cos 2 \theta^{'}_{13} - \sqrt{\Delta m^{2}_{31}} \eta_{ee} \left( m_1 + m_2 - 2 m_3 \right) \left(\cos 2 \theta_{13} + \cos 2 \theta^{'}_{13} \right) \nonumber \\
    &-& \left(2 \Delta m^{2}_{31} \eta^{2}_{ee} + \sqrt{\Delta m^{2}_{31}} \eta_{ee} ( m_1 + m_2 + 2 m_3) + 2V_m \right) \cos 2(\theta_{13} + \theta^{'}_{13}) \nonumber \\
    &+& \cos 2 \theta_{12} \bigg[ \bigg. 3 \Delta m^{2}_{21} + \sqrt{\Delta m^{2}_{31}} \eta_{ee} ( m_2 - m_1) + \Delta m^{2}_{21} \cos 2 \theta^{'}_{13} \nonumber \\
    &+& \sqrt{\Delta m^{2}_{31}} \eta_{ee} ( m_2 - m_1) \left(\cos 2 \theta_{13} + 2 \cos \theta_{13} \cos (\theta_{13}+ 2 \theta^{'}_{13}) \right) \bigg] \bigg. \Bigg\} \Bigg. \nonumber \\
    &+& 2 \left( \left[ 2 \Delta m^{2}_{21} + \sqrt{\Delta m^{2}_{31}} \eta_{ee} ( m_2 - m_1) \right] \cos \theta^{'}_{13} + \sqrt{\Delta m^{2}_{31}} \eta_{ee} ( m_2 - m_1) \cos (2 \theta_{13} + \theta^{'}_{13}) \right) \nonumber \\
    &\times& \sin 2 \theta_{12} \sin 2 \theta^{'}_{12} \Bigg. \Bigg],
\end{eqnarray}

\begin{eqnarray}
    \Delta^{\rm eff}_{31} 
    &=& \frac{1}{4 E} \Bigg[ \Bigg.\left(\Delta m^{2}_{21} + \sqrt{\Delta m^{2}_{31}} \eta_{ee} ( m_2 - m_1) \cos^2 \theta_{13}\right) \cos \theta^{'}_{13} \sin 2 \theta_{12} \sin 2 \theta^{'}_{12} \nonumber \\
    &+& 2 \cos^2 \theta^{'}_{13} \Bigg( \Bigg.\Delta m^{2}_{31} + \cos^2 \theta^{'}_{12} \bigg[ \bigg. -\Delta m^{2}_{21} \sin^2 \theta_{12} - \cos^2 \theta_{13} \Bigl( \Bigl. \Delta m^{2}_{31} \eta^{2}_{ee} + V_m \nonumber \\
    &+& 2 \sqrt{\Delta m^{2}_{31}} \eta_{ee} \left( m_1 \cos^2 \theta_{12} + m_2 \sin^2 \theta_{12} \right) \Bigl) \Bigl.  \bigg] \bigg. +  \left[ \Delta m^{2}_{31} \eta^{2}_{ee} + 2 \sqrt{\Delta m^{2}_{31}} \eta_{ee} m_3 +V_m \right] \sin^2 \theta_{13} \Bigg. \Bigg) \nonumber \\ 
    &+&  \sin \theta^{'}_{13} \bigg\{ \bigg. \frac{1}{2} \sqrt{\Delta m^{2}_{31}} \eta_{ee} \left(m_1 -m_2 \right) \sin 2 \theta^{'}_{12} \sin 2 \theta_{12} \sin 2 \theta_{13} \nonumber \\
    &+& 2 \left( \Delta m^{2}_{21} \sin^2 \theta_{12} + \cos^2 \theta_{13} \left[ \Delta m^{2}_{31} \eta^{2}_{ee} +V_m + 2 \sqrt{\Delta m^{2}_{31}} \eta_{ee} m_2 \sin^2 \theta_{12} \right]   \right) \sin \theta^{'}_{13} \nonumber \\
    &-& 2 \cos^2 \theta^{'}_{12} \left( \Delta m^{2}_{31} + \left[\Delta m^{2}_{31} \eta^{2}_{ee} + 2 \sqrt{\Delta m^{2}_{31}} \eta_{ee} m_3 + V_m \right] \sin^2 \theta_{13}  \right) \sin \theta^{'}_{13}      \bigg\} \bigg.  + \left( 1+ \cos^2 \theta^{'}_{12} \right) \nonumber \\
    &\times& \left(\Delta m^{2}_{31} \eta^{2}_{ee} + \sqrt{\Delta m^{2}_{31}} \eta_{ee} m_3 + V_m + \sqrt{\Delta m^{2}_{31}} \eta_{ee} m_2 \sin^2 \theta_{12} \right) \sin 2 \theta_{13} \sin 2 \theta^{'}_{13} \nonumber \\
    &+& \cos^2 \theta_{12} \Bigg( \Bigg. -2 \Delta m^{2}_{21} \sin^2 \theta^{'}_{12} + \sqrt{\Delta m^{2}_{31}} \eta_{ee} m_1 \bigg( \bigg. 4 \cos^2 \theta_{13} \sin^2 \theta^{'}_{13} \nonumber \\
    &+& \left( 1 + \cos^2 \theta^{'}_{12} \right) \sin 2 \theta_{13} \sin 2 \theta^{'}_{13}   \bigg) \bigg.    \Bigg. \Bigg)   \Bigg. \Bigg].
\end{eqnarray} 

\begin{figure}
    \centering
    \includegraphics[scale=0.48]{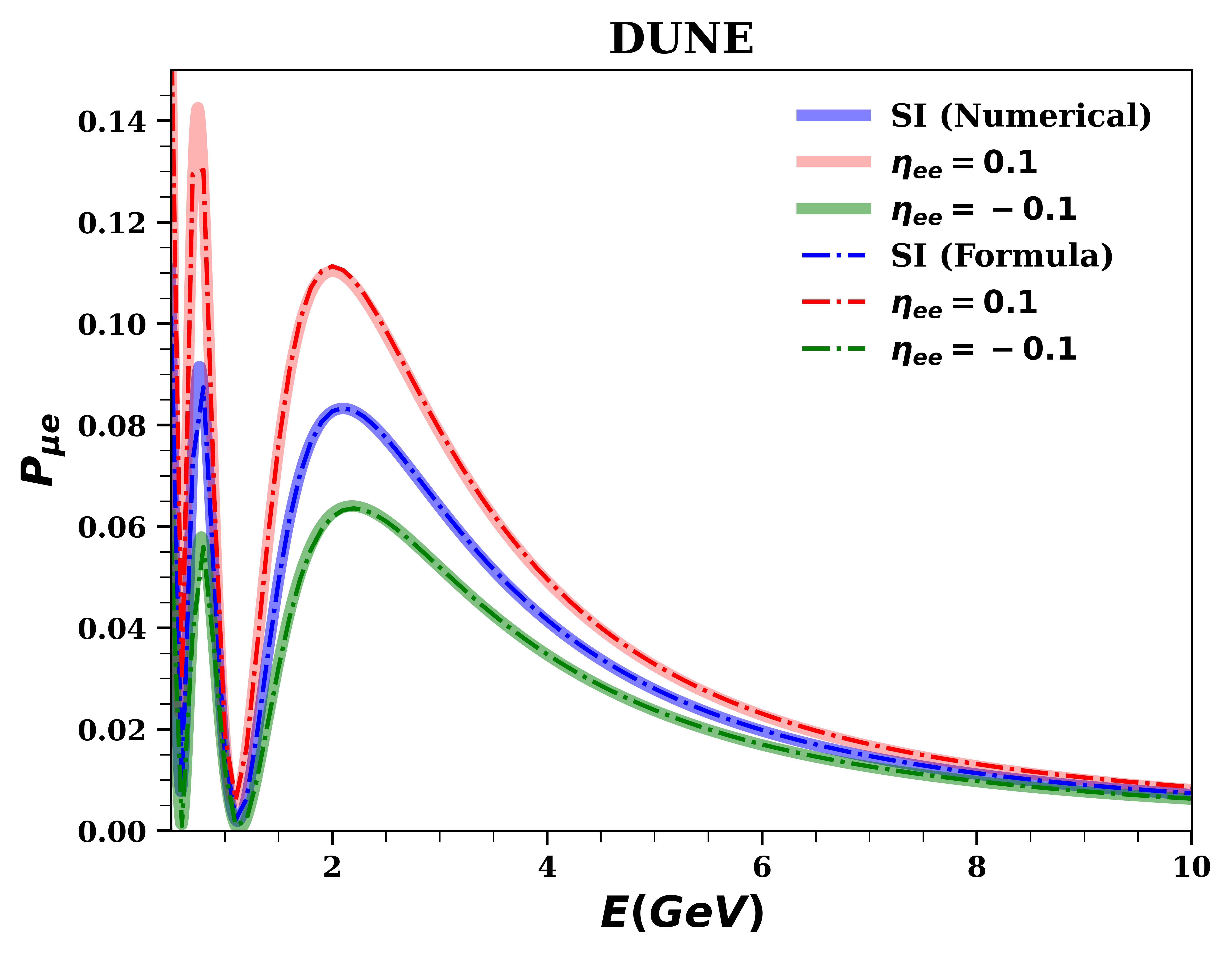}
    \includegraphics[scale=0.48]{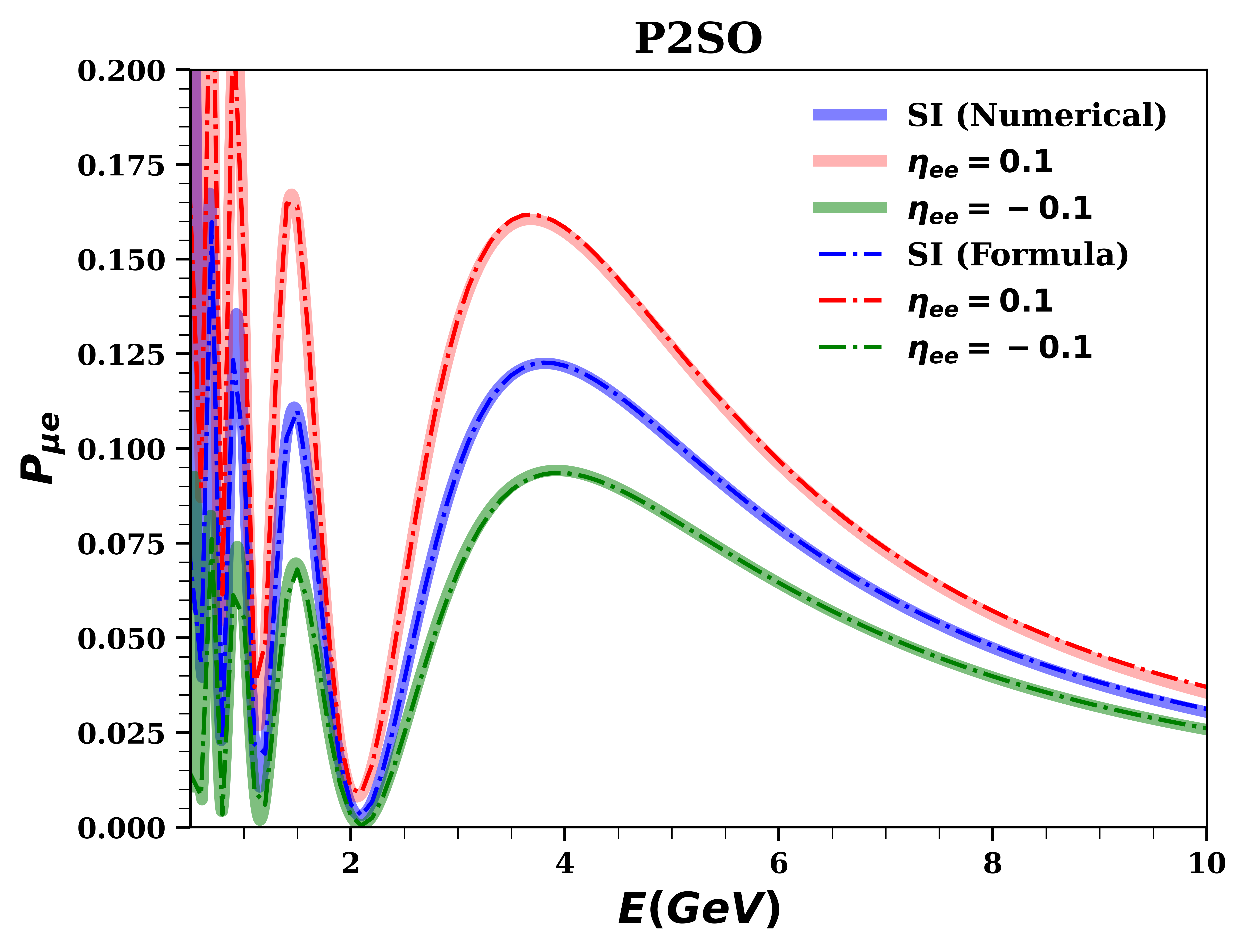}
    \caption{Probability plots showing comparison between the exact numerical estimation and the analytical formula.}
    \label{ana_num}
\end{figure}

In Fig.~\ref{ana_num}, we have shown how our analytical formula matches with the full numerical estimations for DUNE (left panel) and P2SO (right panel). These panels show that our formula matches very will with the numerical estimation.

Though our analytical formula look very complicated from the first glance, some interesting observations can be made from them. In Eq.~\ref{dcp}, if we demand $\sin 2 \theta^{'}_{12} = 0$, the probability becomes independent of $\delta_{\rm CP}$. This condition yields:
\begin{center}
    $\sin 2 \theta^{'}_{12}$ $\sim$ 0  
   $\implies$ $a_{12}$ = 0\;.
\end{center}
This in turn gives
\begin{eqnarray}
    \eta_{ee} &=& \frac{-2 \Delta m^{2}_{21}}{\sqrt{\Delta m^{2}_{31}} (m_2 - m_1) (1+ \cos 2 \theta_{13} - \sin 2 \theta_{13} \tan \theta^{'}_{13})}\;.
\end{eqnarray}
If we assume the contribution to $\theta^{'}_{13}$ to be negligible, then we will get $\eta_{ee}$ $\sim -0.1748 $. This implies the fact that for a value of $\eta_{ee}$ $\sim -0.1748 $, the appearance channel probability becomes independent of $\delta_{\rm CP}$ and the CP sensitivity of an experiment gets lost. This is shown recently in Ref.~\cite{ESSnuSB:2023lbg} in the context of the ESSnuSB experiment. However, this particular value of $\eta_{ee}$ will be ruled out at more than $3 \sigma$ for both P2SO and DUNE if these experiments do not see SNSI.  

\begin{figure}
    \centering
    \includegraphics[scale=0.48]{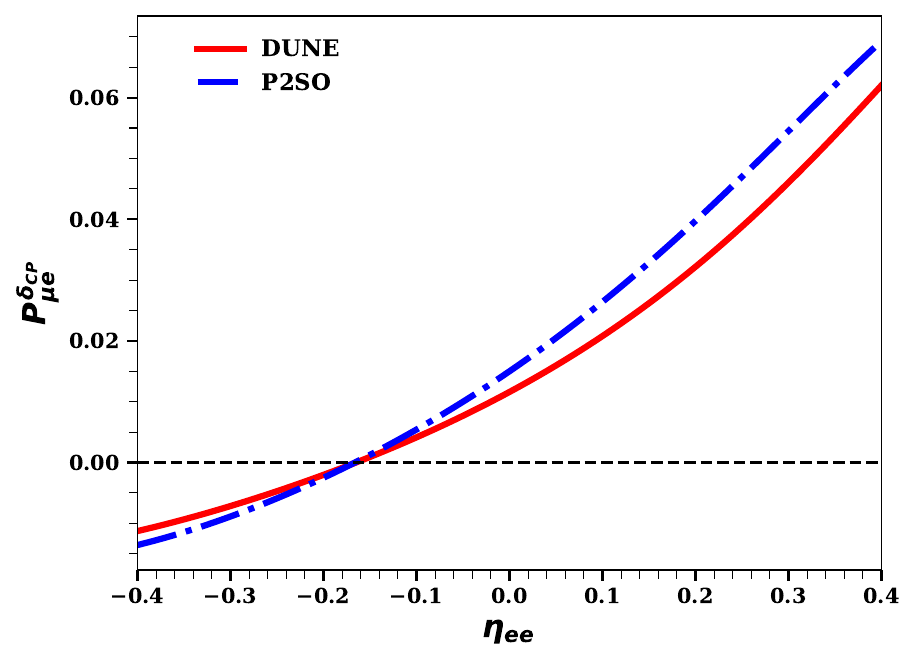}
    \includegraphics[scale=0.48]{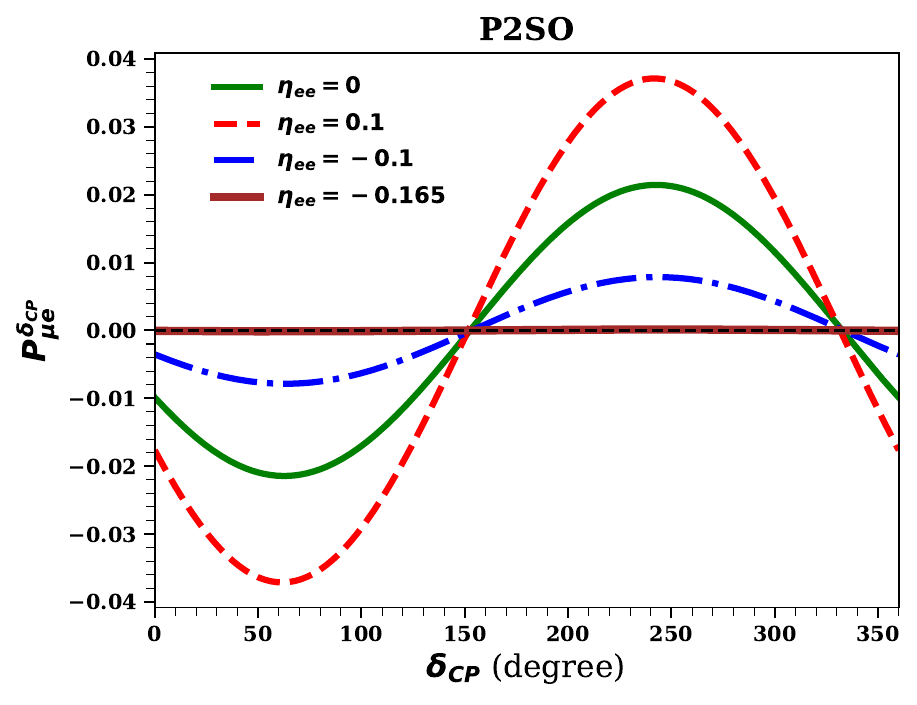}
    \caption{$P^{\delta_{\rm CP}}_{\mu e}$ vs $\eta_{ee}$ (left panel) and $P^{\delta_{\rm CP}}_{\mu e}$ vs $\delta_{\rm CP}$ (right panel).}
    \label{cp_ee}
\end{figure}

For a more clear understanding of the above discussion, in Fig.~\ref{cp_ee}, we have plotted the CP term in the analytical expression i.e.,  Eq.~\ref{dcp}. In these panels, we did not consider $\theta^{'}_{13} = 0$. For DUNE (P2SO) we choose $E = 2.05 (4)$ GeV. From the left panel we see that for $\eta_{ee} = -0.165$, $P^{\delta_{\rm CP}}_{\mu e}$ vanishes for both the experiments. This can be also seen from the brown curve in the right panel where $P^{\delta_{\rm CP}}_{\mu e}$ is plotted against $\delta_{\rm CP}$ for P2SO. Additionally, we observe that for $\delta_{\rm CP} = 155^{\rm o}$ and $335^{\rm o}$,   $P^{\delta_{\rm CP}}_{\mu e}$ vanishes irrespective of the value of $\eta_{ee}$. 

Next let us try to understand why the appearance channel probability is higher (lower) in presence of SNSI as compared to the standard interaction scenario for positive (negative) values of $\eta_{ee}$ (cf. Fig.~\ref{ana_num}) from the analytical expression. For this let us consider a situation when $P^{\delta_{\rm CP}}_{\mu e}$ vanishes i.e., $\delta_{\rm CP} = 155^{\rm o}$ and $335^{\rm o}$. In Eq.~\ref{prob}, one can neglect the $\sin^2 \left(\frac{\Delta^{\rm eff}_{21} L}{2}\right)$ terms and therefore the leading contribution to the probability will come from the non-oscillatory amplitude of $\frac{1}{16} \sin^2 2\left(\theta_{13} + \theta^{'}_{13}\right) \sin^2 \theta_{23}$ as it contains a factor ``7". Now, in the non-oscillatory amplitude, only $\theta_{13}^\prime$ depends on $\eta_{ee}$ via Eq.~\ref{13}. Taking three sample values of $\eta_{ee}$ of $-0.1$, 0 and 0.1, one can numerically estimate that
\begin{equation}
    2 \theta^{'}_{13} (\eta_{ee} = -0.1) = 1.61^{\rm o},~ 2 \theta^{'}_{13} (\eta_{ee} = 0) = 3.56^{\rm o},~ 2 \theta^{'}_{13} (\eta_{ee} = 0.1) = 6.02^{\rm o}\;, \nonumber
\end{equation}
and hence we can conclude that
\begin{equation}
    P_{\mu e} (\eta_{ee} <0) < P_{\mu e} (\eta_{ee} =0) < P_{\mu e} (\eta_{ee} >0).
\end{equation}

\section{CP sensitivity of the diagonal vs off-diagonal SNSI parameters}

\begin{figure} [htb!]
    \centering
    \includegraphics[scale=0.9]{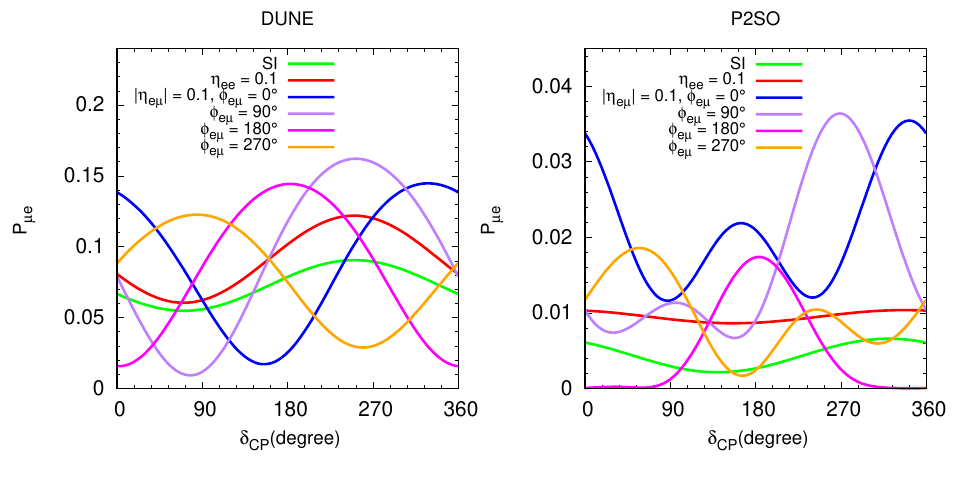}
    \caption{Appearance probability  as a function of \(\delta_{\rm CP}\) for DUNE and P2SO experiment for diagonal and off-diagonal SNSI parameters}
    \label{d_vs_o}
\end{figure}

In section \ref{eff}, we mentioned that the off-diagonal SNSI parameters are more sensitive to the $\delta_{\rm CP}$ as compared to the diagonal SNSI parameters. To show this explicitly, in Fig.~\ref{d_vs_o}, we show the appearance channel probability as a function of $\delta_{\rm CP}$ for both DUNE (left panel) and P2SO (right panel) considering both diagonal and off-diagonal SNSI parameters. From these panels we can clearly see that the variation of the probability with respect to $\delta_{\rm CP}$ is more for the off-diagonal SNSI parameters as compared to the diagonal SNSI parameters.

\bibliography{SNSI.bib}
\bibliographystyle{JHEP}

\end{document}